\documentclass[journal,12pt,draftclsnofoot,onecolumn]{IEEEtran}
\IEEEoverridecommandlockouts
\usepackage{verbatim}
\usepackage{cite}
\usepackage{amsmath,amssymb,amsfonts}
\usepackage{graphicx}
\usepackage{textcomp}
\usepackage{xcolor}
\usepackage{epsfig}
\newtheorem{Remark}{\it Remark}[section]

\newtheorem{Proposition}{\it Proposition}[section]
\newtheorem{Lemma}{\it Lemma}[section]

\newtheorem{Definition}{\it Definition}[section]

\makeatletter
\newcommand{\Rmnum}[1]{\expandafter\@slowromancap\romannumeral #1@}
\makeatother
\usepackage{bm}
\usepackage{multicol}
\usepackage{setspace}
\usepackage{subfigure}
\usepackage{epstopdf}
\usepackage{fancyhdr} 
\usepackage{indentfirst}
\usepackage{enumerate}
\usepackage{stfloats}
\usepackage{bm}
\usepackage{multirow}
\usepackage{threeparttable}
\usepackage{CJK}
\usepackage[justification=centering]{caption}
\usepackage{makecell}
\usepackage{caption}
\usepackage{cases}
\usepackage{amsmath}
\usepackage{algorithm}
\usepackage{algpseudocode}

\usepackage{graphics}
\usepackage{caption}
\usepackage{booktabs}

\definecolor{deepblue}{rgb}{0.6,0,0.4}
\definecolor{red}{rgb}{0.8,0,0}
\definecolor{blue}{rgb}{0.2,0,0.8}

\def\BibTeX{{\rm B\kern-.05em{\sc i\kern-.025em b}\kern-.08em
		T\kern-.1667em\lower.7ex\hbox{E}\kern-.125emX}}
\begin{document}

 \title{ Design and Performance of  Resonant Beam Communications---Part II:  Mobile Scenario}
 \author{\IEEEauthorblockN{Dongxu Li, Yuanming Tian, Chuan Huang, Qingwen Liu, and Shengli Zhou}
 \thanks{
Part of this paper was presented in IEEE ICC 2021.

Dongxu Li and Yuanming Tian are with the Future Network of Intelligence Institute and the School of Science and Engineering, the Chinese University of Hong Kong, Shenzhen 518172, China (emails: dongxuli@link.cuhk.edu.cn; yuanmingtian@link.cuhk.edu.cn).

Chuan Huang is with the School of Science and Engineering and the Future Network of Intelligence Institute, the Chinese University of Hong Kong, Shenzhen 518172, China (email: huangchuan@cuhk.edu.cn).

    Qingwen Liu is with the College of Electronics and Information Engineering, Tongji University, Shanghai 201804, China (email: qliu@tongji.edu.cn).

    Shengli Zhou is with the Department of Electrical and Computer Engineering, University of Connecticut, Storrs, CT 06250 USA (email: shengli.zhou@uconn.edu).
}

}
 \maketitle
 

    

%
\begin{abstract}
This two-part paper focuses on the system design and performance analysis for a point-to-point resonant beam communication (RBCom) system under both the quasi-static and mobile scenarios. Part I of this paper proposes a synchronization-based information transmission scheme and derives the capacity upper and lower bounds for the quasi-static channel case.  In Part II, we address the mobile scenario, where the receiver is in relative motion to the transmitter, and derive a mobile RBCom channel model that jointly considers the Doppler effect, channel variation, and echo interference. With the obtained channel model, we prove that the channel gain of the mobile RBCom decreases as the number of transmitted frames increases,  and thus show that the considered mobile RBCom terminates after the transmitter sends a certain number of frames without frequency compensation. By deriving an upper bound on the number of successfully transmitted frames, we formulate the throughput maximization problem for the considered mobile RBCom system, and solve it via a sequential parametric convex approximation (SPCA) method. Finally, simulation results validate the analysis of our proposed method in some typical scenarios.
  
	\end{abstract}
	\section{Introduction}
With the evolution of modern wireless communications,  optical wireless communication (OWC) \cite{Khalighi} is becoming a promising complement to fulfill the demands for extremely high-speed transmissions, since it has a significantly higher available optical bandwidth\cite{owc1}  compared to radio-frequency (RF) technologies. Moreover, the availability of unregulated spectrum for immediate utilization \cite{Brien} and the widespread presence of light sources\cite{6497926} make OWCs play a significant role in future mobile communication networks.
	
	 Conventional OWC techniques are classified into two categories:  1) omni-directional OWC: This scheme uses light-emitting diode (LED) transmitters to broadcast information bits along all directions and is suitable for the communications with mobile wireless devices\cite{7339420}. However, since the light intensity of LED diminishes dramatically as the distance increases,  this technique is typically limited to the
	 indoor scenarios and cannot achieve very high-speed transmissions\cite{vlc,Zafar};  2) directional OWC: This scheme utilizes the modulated laser beam to realize information transmissions between the transmitter and the receiver at very high data rates. However, the narrow beam of the directional OWC makes it challenging for the transmitter to point to mobile receivers\cite{Kinani,Chowdhury,Hamza}.  To address this issue, an acquisition, tracking, and pointing (ATP) subsystem must be adopted to adaptively track the mobile receiver \cite{ATP_fso}, making it difficult to be equipped in small devices, such as mobile phones. 
	
	Resonant beam communication (RBCom) is a promising OWC technology that simultaneously offers high-data-rate transmission and self-alignment capability for mobile devices\cite{resonantcom}. In RBCom systems, both the transmitter and receiver adopt retroreflectors (e.g., corner cube reflectors \cite{ccr}) to construct a resonant cavity, and the photons in this cavity are amplified by the gain medium to form the stable resonant beam \cite{linford,liu}.  Besides, a unidirectional free-space electro-optic amplitude modulator \cite{van2018model} is utilized to modulate the information bits to the amplitude of the resonant beam, and an optical telescope is placed at the transmitter to decrease the diffraction effect within the resonant cavity \cite{longlaser}. However, there are a few studies that have investigated the impact of mobility on the channel capacity of RBCom. In \cite{mobile_rbc} and \cite{mobile_rbc2}, the authors built a mobile transmission channel model of RBCom based on the self-alignment property of the retroreflectors and then proposed a design scheme for simultaneous energy and information transfer. However,  the impact of the Doppler shift on the mobile RBCom has not been well analyzed.  
	
	In Part II of this paper, we consider a mobile scenario for the point-to-point RBCom system where the receiver is relatively moving to the transmitter. Similar to the quasi-static case studied in Part I of this paper \cite{Dong_part1},
	 the resonant beam propagates cyclically between the transmitter and the receiver.   However, due to the mobility of the receiver, Doppler shift induces a variation in the central frequency of the resonant beam at each reflection round \cite{doppler1}, and the power gain of the gain medium will change according to the carrier frequency \cite{laserbook1}.  To comprehensively analyze the Doppler effect on the communication performance of the mobile RBCom, we build a signal transmission model based on the synchronization-based amplitude modulation method proposed in part I of this paper \cite{Dong_part1}. The main results of Part II of this paper are summarized as follows:
	\begin{enumerate}
	    \item First, we build a link gain model for the mobile RBCom considering both the Doppler shift and the nonlinear gain properties of the gain medium, and reveal a termination behavior of the mobile RBCom after transmitting a certain number of frames. Leveraging this model, we design a new information-bearing scheme for the modulated symbols to simplify the link gain function as a constant within each frame while varying across different frames. After that, we propose an algorithm to compute the upper bound on the number of successfully transmitted frames. Additionally, a Doppler shift compensation method to avoid the termination of mobile RBCom is proposed.
	    \item Then, based on the obtained upper bound on the number of successfully transmitted frames,   the throughput maximization problem for mobile RBCom is formulated and approximately solved by using a sequential parametric convex approximation (SPCA) algorithm: First, we change the design variables and equivalently transform the constraints into the concave-convex form; Then, by substituting the concave parts of the transformed constraints with their first-order Taylor approximations, we obtain a convex approximation of the original problem; Finally, by successively solving a series of the derived convex approximations, we solve the original optimization problem.  
	\end{enumerate} 

   The remainder of this paper is organized as follows. Section II describes the mobile RBCom system model.  Section III builds a link gain model for the mobile RBCom and simplifies the mobile RBCom channel.  Section IV obtains an upper bound on the number of transmitted frames, and solves the throughput maximization problem.   Section V presents the simulation results. Finally, Section VI concludes this paper.

\emph{Notation}: $ \vec{x} $ denotes a vector in the three-dimensional coordinate system, and $ \Vert\vec{x}\Vert $ is the Euclidean norm  of $ \vec{x} $; $ \exp(x) $ is the natural exponent;  $ \ln(x) $ and $ \log_2(x) $   are the natural and  base-$2$ logarithms, respectively;  $ \min\{x, y\} $    and $\max\{x, y\}$ are the minimum  and maximum between two real numbers $ x $ and $ y $, respectively.

	\section{System Model} \label{system_model}
     
   In this section, we investigate a RBCom system where the receiver is in relative motion to the transmitter. Specifically, in the RBCom systems, two retroreflectors are spatially separated at the transmitter and receiver, respectively, to form a resonant cavity due to their self-alignment property \cite{Dong_part1}.  The photons are amplified by the gain medium to oscillate within the resonant cavity to form the resonant beam. Besides, a frame-based transmission scheme is adopted in this RBCom system, i.e., one frame of symbols is transmitted in each reflection round. Moreover, we consider the case that the resonant beam runs $K$ reflection rounds between the transmitter and the receiver.
   
\begin{figure}[htbp]

	\centering
	\includegraphics[width=3.5in]{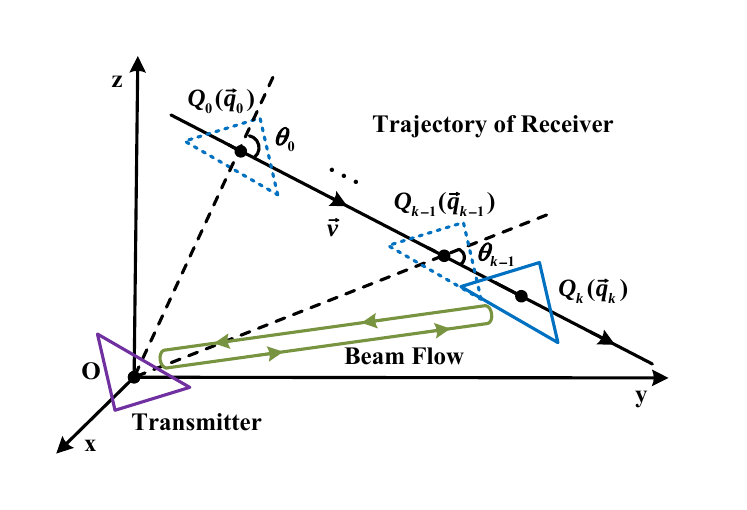}
	\caption{ Mobile RBCom in the $k$-th reflection round.}
	\label{fig_move_1}
    \end{figure}

   As shown in Fig. \ref{fig_move_1}, the transmitter is placed at the original point $ \text{O}$ of the three-dimensional coordinate system, while the receiver, initially located at point $Q_0(\vec{q}_0)$, moves along a fixed direction $\frac{\vec{v}}{\Vert \vec{v} \Vert} \in \mathbb{R}^{3 \times 1}$ with a speed $\Vert \vec{v} \Vert$. Then, during the $k$-th reflection round, $k \leq K$, the receiver moves from point $ Q_{k-1}(\vec{q}_{k-1}) $ to point $ Q_{k}(\vec{q}_k) $, where $\vec{q}_{k-1}\in \mathbb{R}^{3\times 1}$ is the coordinate of point $Q_{k-1}$. Define direction angle $ \theta_{k-1} $ as the angle between vectors $ \vec{q}_{k-1} $ and $ \vec{v} $.  Based on the above setup,   position $ \vec{q}_{k-1} $ and direction angle $ \theta_{k-1} $ 
    in the $k$-th reflection round are derived in the following proposition.
    
 \begin{Proposition} \label{P_qk}
  With given initial position $ \vec{q}_0 $ and velocity $ \vec{v} $ of the receiver, position  $ \vec{q}_k $  and direction angle  $ \theta_{k-1} $ in the $ k $-th reflection round  are recursively computed as 
   \begin{equation}
   	\vec{q}_k = \vec{q}_{k-1} + \frac{2\Vert \vec{q}_{k-1}\Vert\vec{v}}{c},\label{cal_qk}
   \end{equation} 
   \begin{equation}
       	\cos\theta_{k-1} = \frac{\vec{v} \cdot \vec{q}_{k-1} }{\Vert \vec{v}\Vert \Vert \vec{q}_{k-1} \Vert}, 
       	\label{cos_theta}
     \end{equation}
   respectively, $ k = 1, \cdots, K  $.
    \end{Proposition}
    \begin{IEEEproof}
Please see Appendix \ref{ap_qk}.
    \end{IEEEproof}

From \cite{doppler1} and \cite{doppler2}, it is easy to check that the circulating resonant beam undergoes a Doppler shift of $ \frac{2\Vert \vec{v} \Vert f_{k-1}\cos\theta_{k-1}}{c} $ in the $ k $-th reflection round for $k=1,2,\cdots$, with $ f_{k-1} $ being the central frequency of the resonant beam at the beginning of the $ k $-th reflection round and $  c $ being the speed of light. Then, as the resonant beam runs $k$ reflection rounds, the Doppler shift has been accumulated for $ k $ times and thus $f_k$ is recursively calculated as
   \begin{equation}
   	f_k = f_{k-1} -  \frac{2\Vert \vec{v} \Vert f_{k-1}\cos\theta_{k-1}}{c} = f_0 \prod_{i=0}^{k-1}\left(1-\frac{2\Vert \vec{v} \Vert\cos\theta_i}{c}\right),
   	\label{f_k}
   \end{equation} 
   where $ f_0 $ is the initial frequency of the resonant beam, and $\cos\theta_i $ is obtained in Proposition \ref{P_qk}. Moreover, in the $ k $-th reflection round, the distance between the transmitter and the receiver changes from $ \Vert \vec{q}_{k-1}\Vert $ to $ \Vert \vec{q}_{k}\Vert $.  Therefore, duration $ T_k $ of the $ k $-th reflection round is given as
   \begin{equation}
   	T_k = \frac{ \Vert \vec{q}_{k-1} \Vert + \Vert \vec{q}_{k} \Vert  }{c} \approx  \frac{2 \Vert \vec{q}_{k-1} \Vert  }{c}, \label{T_k}
   \end{equation}
where  we approximately treat $ \Vert \vec{q}_{k-1}\Vert  \approx \Vert \vec{q}_{k}\Vert $ since $ \Vert \vec{v} \Vert\ll c $.
\begin{Remark}
	Compared with the quasi-static scenario analyzed in Part I of this paper \cite{Dong_part1}, where each reflection round lasts for equal duration, duration $ T_k $ of the $ k $-th reflection round in the considered mobile scenario varies with $k $ due to the movement of the receiver. Therefore, when applying the frame-based transmission scheme proposed in Part I \cite{Dong_part1} to the considered mobile scenario, i.e., the duration of each frame is equal to that of the corresponding reflection round and the number of transmitted symbols remains constant in each frame, the symbol duration for mobile scenario has to vary with the change of $T_k$. Consequently, to successfully decode the desired information, the sampling frequency at the receiver needs to vary accordingly.
\end{Remark}
     \begin{figure}[htbp]

	\centering
	\includegraphics[width=5.5in]{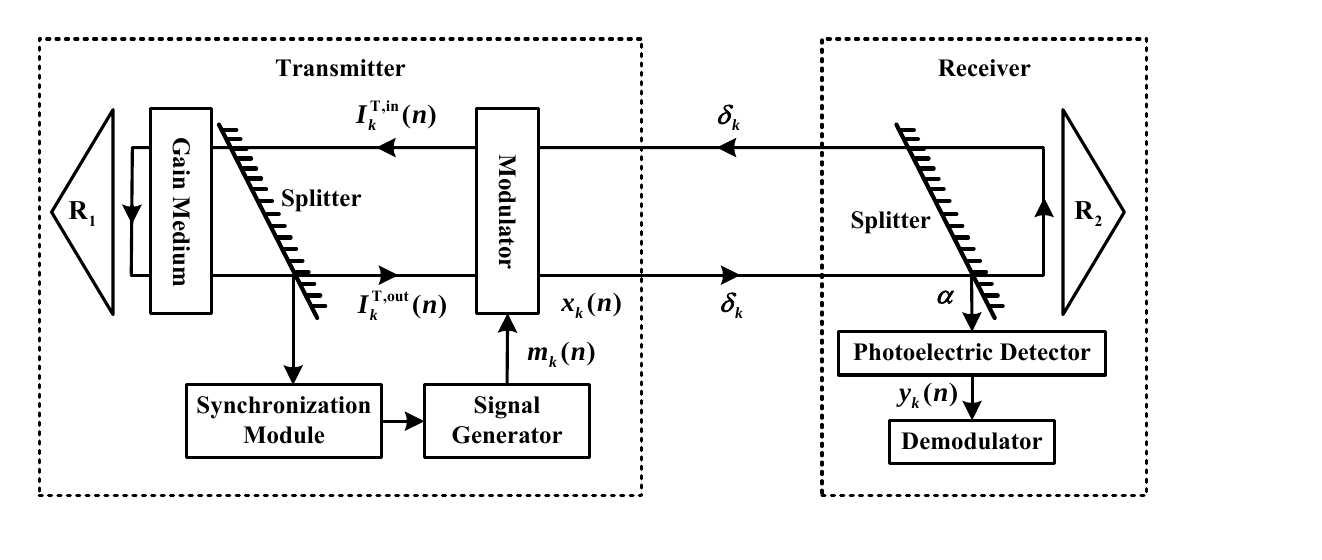}
	\caption{ Transceiver structure of mobile RBCom systems.}
	\label{figure_system}
    \end{figure}   
   \subsection{Transmission at Transmitter}
   We adopt the frame-based transmission scheme \cite{Dong_part1} for the considered mobile RBCom.  Specifically,  
 taking the transmission at frame $k$ as an example (shown in Fig. \ref{figure_system}),  the resonant beam is amplified by the gain medium at the transmitter and then divided into two sub-beams by an optical splitter: One beam with negligible power \cite{Dong_part1} is sent to the synchronization module to achieve symbol synchronization, and the other beam is sent to the amplitude modulator to carry the information bits.  Next, suffering from link loss $ \delta_k $,  the resonant beam at the receiver is also separated into two parts by one optical splitter: The first part with $\alpha $ portion of the total power is detected by a photoelectric detector and then demodulated to recover the transmitted information, and the second part with $1-\alpha$ portion of the total power is reflected by retroreflector $\text{R}_2$ and sent back to the transmitter. Here, we consider the case that the link loss from the receiver to the transmitter is equal to that in the opposite direction of each reflection round.  
 
 Similar to the quasi-static scenario in Part I of this paper \cite{Dong_part1}, 
     we adopt the synchronization-based amplitude modulation approach to transmit multiple frames of symbols under the considered mobile scenario. With this approach, we transmit a synchronization sequence (SS) and then $ N $ transmitted symbols during each frame, where the SS is adopted to achieve symbol-level synchronization across different frames. Meanwhile, the transmitted symbols at frame $ k $ are produced by multiplying the beam reflected back to the transmitter with the newly modulated symbols,  i.e.,
 the $n$-th transmitted symbol $x_k(n)$ at frame $k$  is expressed as
        \begin{equation}
    	x_k(n) =
    	\begin{cases}
    		\sqrt{P_\text{t}}m_k(n), &  k = 1, \\ 
    		\sqrt{h(x_{k-1}^2(n),f_{k-1},\delta_{k-1})}m_k(n), & k = 2,3,\cdots,K, 
    	\end{cases} 
    	n = 1,2,\cdots,N,
    	\label{x_k}
    \end{equation}
	where  $ m_k(n) $, $0 <m_k(n)\leq 1$, is the amplitude-constrained modulated symbol \cite{Dong_part1}, $ P_\text{t} $ is the stable power of the resonant beam before passing through the modulator \cite{Dong_part1},  and $ f_{k-1} $ is computed by \eqref{cal_qk}-\eqref{f_k}. Besides, $ h(x_{k-1}^2(n),f_{k-1},\delta_{k-1}) $ is the link gain as a function of $ x_{k-1}(n)$, $ f_{k-1} $ and $ \delta_{k-1} $, and it represents the instantaneous power of the resonant beam before passing through the modulator at frame $ k $, which jointly characterizes the Doppler shift, amplification of the gain medium and link loss.  The expression for link gain function $ h(\cdot) $ will be derived later. Moreover, the expression for link loss $ \delta_{k-1} $ ($k \geq 2$) has already been obtained in Part I of this paper \cite{Dong_part1}, i.e.,
	   	\begin{equation}
   		\delta_{k-1}=1 - \exp\left( \frac{-2S}{\frac{\lambda^2}{\pi\phi^2} + \pi\phi^2\Vert \vec{q}_{k-1}\Vert^2}\right),
      \label{loss_defi}
   	\end{equation}
	where $ \phi $ is the diffraction angle of the resonant beam \cite{Dong}, $ \lambda $ is the wavelength of the resonant beam, and $ S $ is the effective area of the receiving surface at the receiver.
	
		\subsection{Reception at Receiver} 
	   At the receiver, after being divided by the splitter with ratio $ \alpha $ and then detected at the photoelectric detector,  the $n$-th received symbol $ y_k(n) $ at frame $ k $ is expressed as 
     \begin{align}
      y_k(n)&= \sqrt{\alpha \delta_k}x_k(n) + \nu_k(n) \notag \\ 
       &=\begin{cases}
    		\sqrt{\alpha \delta_k P_\text{t}}m_k(n)+  \nu_k(n) , & k = 1, \\ 
    		\sqrt{\alpha \delta_k h(x_{k-1}^2(n),f_{k-1},\delta_{k-1})}m_k(n)+ \nu_k(n) , & k=2,3,\cdots,K,
    	\end{cases} 
    	 \label{y_k}
     \end{align}       
     where  $ \nu_k(n) $'s are independent and identically distributed (i.i.d.) additive white Gaussian noise (AWGN).
     
\section{ Equivalent Channel Model}

In this section, we introduce the model of the gain medium for the considered mobile scenario and derive the expression for link gain function $h(\cdot)$ in \eqref{x_k}. Then, we prove that the considered mobile RBCom terminates after the transmitter transmits a certain number of frames. Finally, in the $k$-th reflection round before termination, we derive a simplified channel model to approximate $ h(\cdot) $ as a constant within each frame, while it may vary across different frames.
 \subsection{Gain Medium Model and Link Gain Function}
    To derive the expression of $ h(x_k^2(n),f_k,\delta_k) $, we start with developing a model for the gain medium.   In the considered mobile scenario, the power gain for the gain medium is affected by both the intensity and central frequency of the input resonant beam \cite{laserbook1}. Therefore, the power gain in the mobile scenario is defined as the ratio of the output intensity to the input one of the gain medium\cite{laserbook}, i.e.,  
    \begin{equation}
    	G\big(I^{\text{T},\text{in}}_k(n), f_k \big) = \frac{I^{\text{T}, \text{out}}_k(n)}{I^{\text{T}, \text{in}}_k(n)},
    	\label{G_defi}
    \end{equation}
    where $ I^{\text{T},\text{in}}_k(n) $ and $ I^{\text{T}, \text{out}}_k(n) $ are the input and output beam intensities of the gain medium at frame $ k $, respectively. 
     By analyzing the characteristics of the gain medium \cite{laserbook1}, together with our work in Part I of this paper \cite{Dong_part1}, power gain $ G(\cdot) $ defined in \eqref{G_defi} is calculated by the following proposition. 
    	 \begin{Proposition}
	 \label{prop_G}
	  With given $ I^{\text{T}, \text{in}}_k(n) $ and $f_k $, power gain $ G\big(I^{\text{T}, \text{in}}_k(n),f_k\big) $ is uniquely determined by the following equation
     \begin{equation}
     \label{g_calculate}
     	S_\text{g}I^{\text{T}, \text{in}}_k(n) = \frac{\eta P_\text{in}  - \left(1 + \frac{4(f_k-f_0)^2}{\Delta f_{\text{H}}^2} \right)S_\text{g}I_\text{s}(f_0) \cdot  \ln \sqrt{G\big(I^{\text{T}, \text{in}}_k(n),f_k\big)}}{G\big(I^{\text{T}, \text{in}}_k(n),f_k\big)-1},
     \end{equation}
     where $ I_\text{s}(f_0) $ is the saturation intensity at $ f_0 $ \cite{laserbook1}, $ \Delta f_\text{H}$ is the full width at half maximum of the lineshape function \cite{laserbook1} of the gain medium, $ S_{\text{g}} $ is the effective cross-sectional area of the gain medium, $ P_\text{in} $ is the pumping power of the gain medium, and $ \eta $ is the pumping efficiency.
	 \end{Proposition}
\begin{IEEEproof}
	Please see Appendix \ref{ap_G}.
\end{IEEEproof}
     \begin{figure}[htbp]

	\centering
	\includegraphics[width=3.6in]{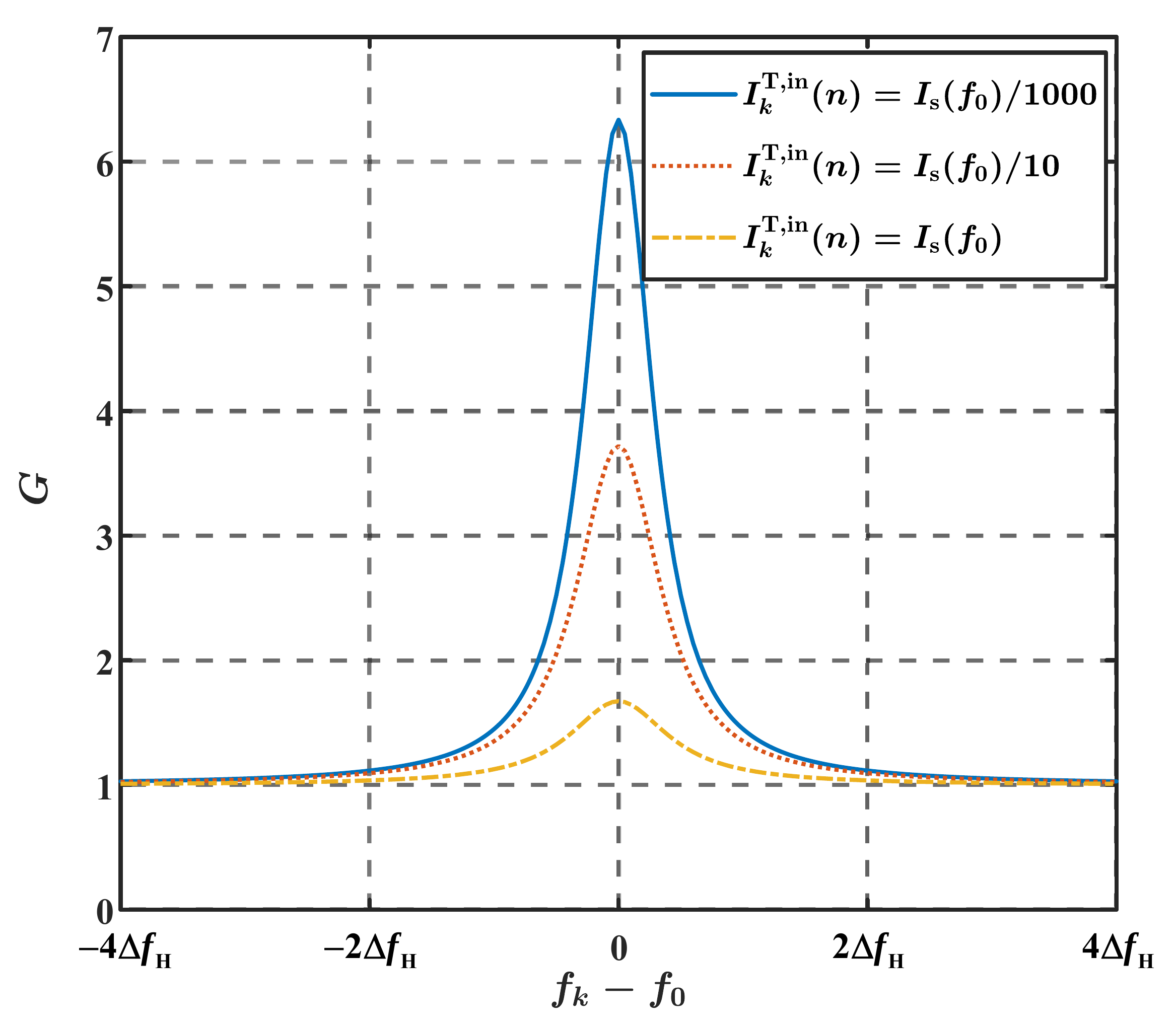}
	\caption{ Power gain $ G\big(I^{\text{T}, \text{in}}_k(n),f_k\big) $ vs.  $ f_k-f_0 $.}
	\label{G_graph}
    \end{figure}  
\begin{Remark} \label{Re_h_k}
	It is noted that parameters $ I_\text{s}(f_0) $, $ \Delta f_\text{H}$, and $ \eta $ in Proposition \ref{prop_G} are dependent on the types of gain mediums and are measured by experiments \cite{laserbook1}.  Moreover, according to \eqref{g_calculate},  we draw the variation of $ G\big(I^{\text{T}, \text{in}}_k(n),f_k\big) $ with respect to $ f_k $ in Fig. \ref{G_graph} by setting $ f_0 = 281.96 $ THz, $ I_\text{s}(f_0)= 1.2 \times 10^7 $ $ \text{W/m}^2 $ \cite{laserbook1}, $ \Delta f_\text{H} = 120 $ GHz\cite{laserbook}, $ S_\text{g} = 12.56$ $ \text{mm}^2$, and $ \eta = 0.7 $, and selecting $ I^{\text{T}, \text{in}}_k(n)$ as $ I_\text{s}(f_0)/1000 $, $ I_\text{s}(f_0)/10$, and $ I_\text{s}(f_0) $, respectively. It is easy to see that $ G(\cdot) $ dramatically decreases as frequency deviation $ |f_k-f_0| $ increases for any fixed $ I^{\text{T}, \text{in}}_k(n)$, and also monotonically decreases as $ I^{\text{T}, \text{in}}_k(n)$ increases. Moreover, the proof of monotonic properties of $ G\big(I^{\text{T}, \text{in}}_k(n),f_k\big) $ can be found in Appendix \ref{ap_G}.
\end{Remark}

 Then, based on link loss model \eqref{loss_defi} and gain medium model \eqref{G_defi}-\eqref{g_calculate}, link gain function $ h(x_k^2(n),f_k,\delta_k)$  is expressed as
   	 \begin{equation}
   	 \label{func_h}
   	 	h(x_k^2(n), f_k, \delta_k ) =  (1-\alpha)\delta_k^2x_k^2(n) G\left(\frac{(1-\alpha)\delta_k^2x_k^2(n) }{S_{\text{g}} },f_k\right),
   	 \end{equation}
   	 where  $ \frac{(1-\alpha)\delta_k^2x_k^2(n) }{S_{\text{g}} }$ is the expression for input intensity $ I^{\text{T}, \text{in}}_k(n) $  of the gain medium at frame $ k $.  Moreover,  link gain function $ h(\cdot)$ has the following decreasing property.

 \begin{Proposition}
  \label{prop_h_k}
  	In the considered mobile scenario as shown in Fig. \ref{fig_move_1}, link gain function $ h(x_{k-1}^2(n),f_{k-1},\delta_{k-1}) $ in \eqref{x_k} decreases to $0$  approximately exponentially, as $ k $ becomes sufficiently large.
  \end{Proposition} 
  \begin{IEEEproof}
      Please see Appendix \ref{ap_prop_h_k}.
  	  \end{IEEEproof}
  \begin{Remark} From Proposition \ref{prop_h_k}, we observe that:
  \begin{enumerate}
  	\item As discussed in Appendix \ref{ap_prop_h_k}, when $ k $ is sufficiently large, frequency deviation $|f_k-f_0|$ also becomes quite large due to the accumulation of Doppler shift given in \eqref{f_k}. Correspondingly, power gain $ G(\cdot) $ depicted in Fig. \ref{G_graph} is then close to $1$, showing that the gain medium produces almost no gain to the passing beam. Then, by \eqref{ap_h_decrease},  $ h(x_{k-1}^2(n),f_{k-1},\delta_{k-1}) $ is proved to be almost exponentially decreasing to $ 0 $ with respect to $ k $ when $ k $ is sufficiently large. 
  	\item  Due to the exponentially decreasing property of $ h(x_{k-1}^2(n),f_{k-1},\delta_{k-1}) $, it is easy to see that the power of the received symbol $ y_k(n) $ in \eqref{y_k} also decreases to $0$  approximately exponentially, as $ k $ becomes sufficiently large. Then, the considered mobile RBCom terminates after the transmitter sends a certain number of frames, and thus a compensation method to combat the Doppler shift and termination of the mobile RBCom is needed, which will be discussed in the next section.
  \end{enumerate}

  \end{Remark}

  \subsection{Simplified Channel Model}
  Similar to the quasi-static scenario in Part I of this paper \cite{Dong_part1},   the signal model in \eqref{y_k} reveals that mobile RBCom suffers from echo interference, i.e., transmitted symbol $ x_{k-1}(n) $ running cyclicly between the transmitter and the receiver interferes the subsequently transmitted symbol $ x_k(n) $. 
   	  
  To address the echo interference issue when the RBCom transmits multiple frames of symbols, we propose a new design scheme for the modulated symbol $ m_k(n) $ in \eqref{x_k}. Similar to Part I of this paper \cite{Dong_part1}, modulated symbol $ m_k(n) $  is designed as
   \begin{equation}
   	m_k(n) = w_k(n)s_k(n),
   	\label{m_k}
   \end{equation}
   where  $ s_k(n) $,  $0 \leq \mu_k \leq s_k(n) \leq 1 $, is the amplitude-constrained i.i.d. information symbol in frame $ k $. Besides, 
   considering the decreasing property of $ h(x_{k-1}^2(n),f_{k-1},\delta_{k-1})$ proved in Proposition \ref{prop_h_k}, $ w_k(n) \in(0,1] $ is adopted to normalize $ \sqrt{h(x_{k-1}^2(n),f_{k-1},\delta_{k-1})}$ to a constant within frame $k$, i.e.,  
      \begin{equation}
	\label{w_k}
	\begin{cases}
	     \sqrt{P_\text{t}}w_k(n) = A_k, & k=1, \\   
		\sqrt{h(x_{k-1}^2(n),f_{k-1},\delta_{k-1})}w_k(n) = A_k, & k=2,3,\cdots,K,
	\end{cases}   	     
   \end{equation}
    where $ A_k $ is the transformed channel coefficient and may vary across different $ k $. Since $w_k(n) \in(0,1]$, it is easy to check that \eqref{w_k} does not always hold, and the sufficient and necessary condition for \eqref{w_k} to be valid is obtained as follows.
   \begin{Proposition}
  Based on the design scheme \eqref{m_k}-\eqref{w_k} for the modulated symbol $ m_k(n) $, equation \eqref{w_k} holds  if and only if $ A_k $ satisfies 
  \begin{equation} \label{Ak_range}
  	 0\leq  A_k \leq \begin{cases}
   	     \sqrt{P_\text{t}}, &k = 1,  \\
   	   \sqrt{h(A_{k-1}^2\mu_{k-1}^2 ,f_{k-1},\delta_{k-1})}  &k =2,3,\cdots,K.
   	   \end{cases}
  \end{equation}
   	\label{prop_wk}
   \end{Proposition}
   \begin{IEEEproof}
   Please see Appendix \ref{ap_P_wk}.
    \end{IEEEproof}
    
   Then, by using \eqref{m_k} and \eqref{w_k}, the channel input and output model in \eqref{y_k} for the mobile RBCom is simplified as 
   \begin{equation}
   	y_k(n) = \sqrt{\alpha\delta_k}A_ks_k(n) + v_k(n),
    	  \label{AWGN_model} 
   \end{equation}
   for all $k =1, 2, \cdots,K$, $\ n=1,2,\cdots,N$, and $ s_k(n)\in[\mu_k, 1]$, where the channel coefficient of the mobile RBCom is now simplified as a constant within each frame, while may vary across different frames. 
   
 \begin{Remark}
 From the simplified channel model in \eqref{AWGN_model}, we observe that:
 \begin{enumerate}
 	\item Due to the exponentially decreasing property of $ h(x_{k-1}^2(n),f_{k-1},\delta_{k-1}) $ and bounded coefficient $ w_k(n) $,    it is easy to demonstrate from \eqref{w_k} that the transformed channel coefficient  $A_k $ also decreases approximately exponentially to $0$ when $ k $ is sufficiently large. Therefore, there exists an upper bound on the number of successfully transmitted frames in the considered mobile RBCom.
 	\item In the quasi-static scenario discussed in Part I of this paper \cite{Dong_part1}, the RBCom channel is simplified as a group of parallel amplitude-constrained AWGN channels to effectively eliminate the echo interference. However, in the considered mobile RBCom,  the echo interference issue persists, and link gain function $ h(\cdot) $ in \eqref{x_k} exhibits a decaying property, as proved in Proposition \ref{prop_h_k}. To deal with these issues, the transformed channel coefficient $A_k$ is designed to be a constant within each frame, while varying across different frames. 
 \end{enumerate}          
 \end{Remark}

 \section{ Performance Optimization} 
 
 Based on the simplified channel model derived in \eqref{AWGN_model}, this section discusses the upper bound on the number of successfully transmitted frames before the mobile RBCom terminates. Moreover, we propose a Doppler shift compensation method to avoid the termination of the mobile RBCom. Then, utilizing the obtained upper bound, we formulate the throughput maximization problem and adopt an SPCA method \cite{Beck} to solve this problem.
  
\subsection{Upper bound on the number of successfully transmitted frames}

As previously stated in \eqref{x_k}, our approach employs a frame-based transmission scheme to convey $ N $ symbols in each frame, with the same duration as the corresponding reflection round. Then, based on the Nyquist theorem \cite{por2019nyquist}, together with duration $ T_k $ of the $ k $-th reflection round derived in \eqref{T_k}, the required channel bandwidth $ B_k $ to send frame $ k $ is given as 
   \begin{equation} \label{B_k}
   	B_k = \frac{N}{2T_k}=\frac{Nc}{4\Vert \vec{q}_{k-1} \Vert}.
   \end{equation}
  Then, for the simplified channel model in \eqref{AWGN_model}, where channel coefficient $A_k$ remains constant and information symbols $s_k(n)$'s are i.i.d. and amplitude-constrained within frame $k$, a tight approximation on the channel capacity  at frame $k$ is given as \cite{Dong_part1,McKellips,Thangaraj}
  
      \begin{equation}
     C_{\text{approx}}( P_{k}, B_k) = \begin{cases}
		  B_k \log_2 \left( 1 + \sqrt{\frac{2P_{k}}{\pi e n_0B_k }}  \right), &\mathrm{if} \ \frac{P_{k}}{n_0B_k}  >  \frac{ 8  }{ \pi e \left( 1 - \frac{2}{\pi e} \right)^2 },  \\
		   \frac{B_k}{2} \log_2\left(1+\frac{P_{k}}{n_0B_k } \right),  &\mathrm{otherwise},
	\end{cases}
     \label{appro_C}
    \end{equation}
    with 
   \begin{equation}
   	  P_{k} = \frac{(1-\mu_k)^2\alpha\delta_kA_k^2}{4}
    	\label{P_peak}
   \end{equation}
   being the peak received signal power at frame $ k $ \cite{smith}. Besides, $ n_0 $ is the one-side power spectrum density (PSD) of the AWGN noise, and $e$ is Euler's number.   Moreover, $ C_{\text{approx}}( P_{k}, B_k) $ has the following properties.

 \begin{Lemma}
 \label{le_decrease_C}
 	$ C_{\text{approx}}( P_{k}, B_k) $  goes to $0$,  when  $ k $ is sufficiently large.
 \end{Lemma} 
 \begin{IEEEproof}
 	  From the simplified channel model in \eqref{AWGN_model}, we have shown that $ A_k $  decreases approximately exponentially to $0$ when $ k $ is sufficiently large. Then, by \eqref{P_peak} and the fact of $ 0 \leq \delta_k$ and $\mu_k \leq 1 $ obtained from \eqref{loss_defi} and \eqref{m_k},  it is easy to see that the peak received signal power $ P_k $ also goes to $0$ approximately exponentially as $ k $ is sufficiently large. Finally, Lemma \ref{le_decrease_C} is obtained by checking \eqref{appro_C}.
 \end{IEEEproof} 
 
 Now, we are ready to rigorously define the number of successfully transmitted frames.
 \begin{Definition} 
 \label{def_k}
 The considered mobile RBCom can only successfully transmit $ k_0 $ frames if $ C_{\text{approx}}( P_{k}, B_k) \geq C_{\text{th}} $ for $ k \leq k_0 $ and $  C_{\text{approx}}( P_{k_0+1}, B_{k_0+1}) < C_{\text{th}} $, where $ C_{\text{th}}  $  is a given minimum data rate.
 \end{Definition} 
 \begin{Proposition}
\label{prop_K0}
	 $ K_0 $ is an upper bound on the number  $ k_0 $ of the successfully transmitted frames if $ C_{\text{approx}}(\cdot)$ satisfies $  C_{\text{approx}}(\frac{\alpha\delta_{K_0+1} A_{K_0+1}^2}{4}, B_{K_0+1}) < C_{\text{th}} $, with $ A_{K_0+1}^2 $ being recursively computed as
	 \begin{equation} \label{Ak_up}
	 	   A_{K_0+1}^2=  \begin{cases}
   	     P_\text{t}, &K_0 = 0,  \\
   	   h(A_{K_0}^2,f_{K_0},\delta_{K_0}),   &K_0 =1,2,\cdots,
   \end{cases}
	 \end{equation}
	 and $ \delta_{K_0+1} $ and $ B_{K_0+1} $ being obtained by \eqref{loss_defi} and \eqref{B_k}, respectively.
\end{Proposition} 
     
%

\begin{IEEEproof}
Please see Appendix \ref{ap_prop_K0}.
\end{IEEEproof} 
 Thereby, we propose an iterative algorithm summarized as Algorithm \ref{al_K_0} to compute upper bound $ K_0 $. The main procedure of this algorithm is described as follows.

As proved in part I of this paper \cite{Dong_part1}, for a fixed splitting ratio $ \alpha $, pumping power $ P_\text{in} $ must be larger than a threshold power $ P_\text{th} $ to generate the resonant beam, which  is given as \cite{Dong_part1}
\begin{equation}
	P_\text{th} = -\frac{(2\ln \delta_0 + \ln (1-\alpha))I_\text{s}(f_0)S_\text{g}
  	  }{2 \eta }.
  	  \label{P_th}
\end{equation}
 Therefore, in Algorithm \ref{al_K_0}, we first compute  $ P_\text{th} $ by \eqref{P_th} and then   set  $ P_\text{in} $ to satisfy  $ P_\text{in} > P_\text{th} $. Then, 
   note that the resonant beam is considered to be stable with power $ P_\text{t} $ at the transmitter before the starting of communications, which implies (by \eqref{g_calculate} and \eqref{func_h}) 
  	  \begin{equation}
  	  	h(P_\text{t}, f_0, \delta_0 ) =P_\text{t}. 
  	  	\label{Pt}
  	  \end{equation}
  	  Besides,  as proved in Lemma \ref{ap_Le_h_property} in Appendix \ref{ap_prop_h_k}, function $ \frac{h(P_\text{t}, f_0, \delta_0 )}{P_\text{t}} $ monotonically increases with respect to $ P_\text{t} $. Therefore, stable power $ P_\text{t} $  can be approximately computed by \eqref{Pt} via bisection search.
 After that, we proceed to iteratively compute $ f_{K_0+1} $,  $ \delta_{K_0+1} $, $ A_{K_0+1} $, and $B_{K_0+1} $ by using \eqref{f_k}, \eqref{loss_defi}, \eqref{Ak_range}, and \eqref{B_k}, respectively.  Finally, $K_0 $ becomes an upper bound on the number of successfully transmitted frames if the condition given in Proposition \ref{prop_K0} is satisfied. Additionally, we denote $ T_\text{up}^{K_0} $ as the upper bound on the communication time of the mobile RBCom system, which is computed by summing the durations of transmitted $ K_0 $ frames, i.e., 
 \begin{equation}
 	T_\text{up}^{K_0} = \sum_{k=1}^{K_0}T_k. \label{T_up}
 \end{equation}
 
\begin{Remark}
	 In the above analysis, we have discussed the termination of the considered mobile RBCom and revealed that the main reason for this termination is the accumulation of frequency deviation caused by Doppler shift, as described by \eqref{f_k} and \eqref{ap_h_decrease}. In order to prevent this termination and increase the communication time in \eqref{T_up}, we propose a Doppler shift compensation method as follows: First, an electronic feedback loop is added to the transmitter to detect the frequency of the resonant beam in real time. Then, we introduce an extra frequency conversion device, e.g., inorganic crystals \cite{bordui1993inorganic}, to the resonant cavity. When the frequency exhibits a large deviation from the initial frequency $ f_0 $, the frequency conversion device becomes operational and adjusts the frequency of the resonant beam back to its initial value of $ f_0 $ \cite{bordui1993inorganic}.   Through the utilization of this method, the frequency deviation $ |f_k - f_0| $ in \eqref{ap_h_decrease} is confined to a reasonable range, thereby preventing the link gain function $ h(\cdot) $ from decreasing to $0$. This, in turn, ensures that the channel capacity $ C_{\text{approx}}( P_{k}, B_k) $ does not diminish to $0$ and increases the communication time of the considered mobile RBCom.     
\end{Remark} 
   \begin{algorithm}[htbp]
\label{alg_1}
	\caption{Compute the upper bound $ K_0 $ on the number of successfully transmitted frames. \label{al_K_0} }
	\label{123} 
	\begin{algorithmic}[1]
	\Require  $\phi$, $ \eta $, $ B_1 $, $ S $, $ f_0 $, $ I_\text{s}(f_0)$, $ S_{\text{g}} $, $ n_0 $, $ \alpha $, $ \vec{v} $, $ \vec{q}_0 $, and $ C_\text{th} $.
	\Ensure $ K_0 $.
    \State Compute threshold power $ P_\text{th} $ by \eqref{P_th}; 
    \State Set pumping power $ P_\text{in} $ satisfying $ P_\text{in} > P_\text{th} $; 
    \State Set $ K_0 = 0 $;
    \State  Compute $ P_\text{t} $ by \eqref{Pt} via bisection search, and let $ A_1^2 = P_\text{t}$;
    \State Compute  $ \delta_1 $ and $ N $ by \eqref{loss_defi} and \eqref{B_k}, respectively;
    \State \textbf{While}  $ C_{\text{approx}}(\frac{\alpha\delta_{K_0+1} A_{K_0+1}^2}{4}, B_{K_0+1}) > C_{\text{th}} $ \textbf{do} 
      \State \quad \ \ Let $K_0 = K_0 +1$;
    \State \quad \ \ Compute $ f_{K_0} $ by \eqref{f_k};
    \State \quad \ \ Update  $ A_{K_0+1}^2 \leftarrow  h(A_{K_0}^2,f_{K_0},\delta_{K_0})$ by \eqref{Ak_range}; 
    \State \quad \ \ Compute  $ \delta_{K_0+1} $ and $ B_{K_0+1} $ by \eqref{loss_defi} and \eqref{B_k}, respectively; 
    \State \textbf{End while}
    \State Compute upper bound $ T_\text{up}^{K_0} $ of communication time by \eqref{T_up}.
	\end{algorithmic}
\end{algorithm}   
   \subsection{Throughput Maximization Problem}
  Considering the finite transmission process discussed in the previous subsection, the throughput maximization problem for the considered mobile RBCom is formulated as
   \begin{eqnarray}
     	\text{($ \text{P}_1 $)}  &\max\limits_{\left\{ \substack{  \mu_1, \cdots,\mu_{K_0} \\ A_1,\cdots,A_{K_0} }\right\}} & \sum_{k=1}^{K_0} T_kC_{\text{approx}}(P_{k},B_k)\label{obj}   \\
  &\text{s.t.} 
  & \mu_k \in [0,1],  \forall k \in \{1,\cdots, K_0\},    \eqref{Ak_range}, \eqref{P_peak}, \label{P1_constraint} \\
  &  &0\leq A_k\leq \sqrt{\frac{P_\text{r,max}}{\alpha\delta_k}}, \quad k =1,\cdots,K_0,
  \label{P1_constraint2}
  \end{eqnarray}
  where the objective function in \eqref{obj} is the throughput over the $ K_0 $ transmitted frames, $ T_k $ is the duration of the $ k $-th frame and it is calculated by \eqref{T_k}, and $B_k $ is computed by \eqref{B_k}.  Besides, $ P_k $ in \eqref{obj} is computed by $\mu_k$ and $A_k$ through \eqref{P_peak}, $ \mu_k \in[0,1] $ is obained by \eqref{m_k},  and $ A_k $ in \eqref{P_peak} is bounded by \eqref{Ak_range} and \eqref{P1_constraint2}, for $ k=1,\cdots, K_0 $. Here, \eqref{P1_constraint2} is obtained by the maximum received signal power constraint \cite{Dong_part1} and \eqref{AWGN_model}. 
   To make Problem \text{($ \text{P}_1 $)} tractable, it is equivalently written as
 \begin{eqnarray}
   	\text{($ \text{P}_2 $)}  &\max\limits_{ \left\{ \substack{  \mu_1, \cdots,\mu_{K_0} \\ A_1,\cdots,A_{K_0} \\ P_1,\cdots,P_{K_0} }\right\}} &\sum_{k=1}^{K_0}T_kC_{\text{approx}}(P_{k},B_k)  \\
  	&\text{s.t.}  &0 \leq P_k \leq  \frac{\alpha\delta_k (1-\mu_k)^2A_k^2}{4}, \quad k =1,\cdots,K_0,  \label{Pk_inequal} \\
  	&\  &0\leq A_1\leq \sqrt{\min \left\{P_\text{t},\frac{P_\text{r,max}}{\alpha\delta_1}\right\}}, \label{A1_inequal}
  	 \\
  	&\ &A_k^2 \leq h(A_{k-1}^2\mu_{k-1}^2 ,f_{k-1},\delta_{k-1}) , \quad k =2,\cdots,K_0,\label{Ak_inequal1}  \\
  	&\    &0\leq A_k\leq \sqrt{\frac{P_\text{r,max}}{\alpha\delta_k}}, \quad k =2,\cdots,K_0,  \label{Ak_inequal2}
  	\\
  &\  &\mu_k \in [0,1],  \forall k \in \{1,\cdots, K_0\},    \label{mu_inequal}
   \end{eqnarray}
 where \eqref{Pk_inequal}  is  obtained from equality constraint \eqref{P_peak} since $ C_\text{approx} $ in \eqref{appro_C} monotonically increases with respect to $ P_k $. Besides,  \eqref{A1_inequal}-\eqref{Ak_inequal2}  are directly derived from \eqref{Ak_range} and \eqref{P1_constraint2}, and \eqref{mu_inequal} is the feasible region of $\mu_k$ given in \eqref{P1_constraint}.
   
   \begin{Remark}
  It is easy to see that the objective function of Problem ($ \text{P}_2 $) is concave with respect to variables $ P_{1}, P_{2},$ $ \cdots, P_{K_0}$ \cite{boyd2004}. However, the non-convex property of constraints \eqref{Pk_inequal} and \eqref{Ak_inequal1} brings a challenge in solving this problem\cite{boyd2004}. To address this challenge, we 
    adopt an iterative algorithm named SPCA to tackle this problem, which can derive an approximate solution for Problem ($ \text{P}_2 $). The convergence of the SPCA algorithm was discussed in \cite{Beck}.
   \end{Remark}

 \subsection{Approximation and Solution to Problem ($ \text{P}_2 $)}
In this subsection, we employ an SPCA method \cite{Beck} to tackle Problem ($ \text{P}_2 $). The fundamental idea of this method is to replace each non-convex constraint with its convex upper bound and then progressively optimize the convex approximation of Problem ($ \text{P}_2 $). We outline the main steps of the SPCA method as follows:
\begin{enumerate}
    \item Approximate link gain function $ h(A_{k-1}^2\mu_{k-1}^2,f_{k-1},\delta_{k-1}) $ in \eqref{Ak_inequal1}  by a linear function of $ A_{k-1}^2\mu_{k-1}^2 $, and then obtain Problem ($ \text{P}_3 $) from Problem ($ \text{P}_2 $);
	\item Change variables $ \mu_1, \cdots, \mu_{K_0} $ and $ A_1, \cdots, A_{K_0} $ in Problem ($ \text{P}_3 $), and then equivalently transform inequalities \eqref{Pk_inequal} and \eqref{Ak_upper_ineq} into the concave-convex form;
	\item Transform the concave-convex constraints into convex forms, and then derive and solve a sequence of convex approximations for Problem ($ \text{P}_2 $).
\end{enumerate}

\subsubsection{Approximation of   $ h(A_{k-1}^2\mu_{k-1}^2 ,f_{k-1},\delta_{k-1})$}

From the proof of Proposition \ref{prop_G} and \eqref{func_h}, it is easy to show
\begin{equation}
	h(A_{k-1}^2\mu_{k-1}^2 ,f_{k-1},\delta_{k-1}) \leq e^{2g_0(f_{k-1})l}(1-\alpha)\delta_{k-1}^2A_{k-1}^2\mu_{k-1}^2,
\end{equation}
where $g_0(f_{k-1}) $ is given in \eqref{ap_g_fk}. Noting that the expression for $h(\cdot)$ given by \eqref{g_calculate} and \eqref{func_h} is quite complicated,
for simplicity, we approximately treat function  $ h(A_{k-1}^2\mu_{k-1}^2 ,f_{k-1},\delta_{k-1})$  as 
\begin{equation}
	h(A_{k-1}^2\mu_{k-1}^2 ,f_{k-1},\delta_{k-1}) \approx e^{2g_0(f_{k-1})l}(1-\alpha)\delta_{k-1}^2A_{k-1}^2\mu_{k-1}^2.
	\label{approx_h}
\end{equation} 
Then, Problem ($ \text{P}_2 $) can be approximated as 
 \begin{align}
   	\text{($ \text{P}_3 $)}    \max\limits_{ \left\{\substack{  \mu_1, \cdots,\mu_{K_0} \\ A_1,\cdots,A_{K_0} \\ P_1,\cdots,P_{K_0} }\right\}} &\sum_{k=1}^{K}T_kC_{\text{approx}}(P_{k},B_k)  \\
  	\text{s.t.} \quad \; \; \;
  	&A_k^2 \leq e^{2g_0(f_{k-1})l}(1-\alpha)\delta_{k-1}^2A_{k-1}^2\mu_{k-1}^2  , \quad k =2,\cdots,K_0,  \label{Ak_upper_ineq} \\
    &\eqref{Pk_inequal}, \eqref{A1_inequal}, \eqref{Ak_inequal2}, \eqref{mu_inequal},
   \end{align}
   where \eqref{Ak_upper_ineq} is obtained by  substituting $h(\cdot)$ in \eqref{Ak_inequal1} with \eqref{approx_h}.
   
   \subsubsection{Changing variables}
   
  We use exponential functions to replace optimization variables $  \mu_1, \cdots,\mu_{K_0} $ and $A_1,\cdots,A_{K_0} $, i.e., define
   \begin{equation}
   	\mu_k = e^{ \nu_k}, \ A_k = e^{\frac{1}{2}a_k}, \quad k = 1, \cdots, K_0.  \label{new_vari_eq}
   \end{equation}
  Besides, the SPCA method \cite{Beck} requires the feasible sets of new optimization variables $ \nu_k$ and $ a_k $ to be compact, i.e., their feasible sets must be closed and bounded. However,  it is easy to see $ \nu_k$ and $ a_k\to-\infty $ in the case of $ \mu_k = A_k = 0 $ in \eqref{Ak_inequal2} and \eqref{mu_inequal}. To ensure the SPCA method remains feasible,  we consider the case that  $ \nu_k$ and $ a_k $ are both greater than or equal to a sufficiently small real number $\beta $, for $ k = 1, \cdots, K_0 $. With these transformations, we convert Problem ($ \text{P}_3 $) into
\begin{align}
   	( \text{P}_4 )    \max\limits_{ \left\{\substack{ \nu_1, \cdots,\nu_{K_0} \\ a_1,\cdots,a_{K_0} \\ P_1,\cdots,P_{K_0} }\right\}} &\sum_{k=1}^{K_0}T_kC_{\text{approx}}(P_{k},B_k)  \\
  	\text{s.t.} \quad \; \; \; 	&- e^{2g_0(f_{k-1})l}(1-\alpha)\delta_{k-1}^2e^{2\nu_{k-1}+a_{k-1}}  + e^{a_k} \leq 0 , \quad k =2,\cdots,K_0,\label{non_convex1}  \\
  	   &\frac{\alpha\delta_{k-1}}{2}e^{\nu_k+a_k}-\frac{\alpha\delta_{k-1}}{4}e^{2\nu_k+a_k}-\frac{\alpha\delta_{k-1}}{4}e^{a_k} +P_k  \leq 0   ,  k =1,\cdots,K_0, \label{non_convex2} \\
  	   &\beta  \leq a_1\leq \min\left\{ \ln( P_\text{t}),\ln\left(\frac{P_\text{r,max}}{\alpha\delta_1}\right)\right\}, \label{a1_inequal} \\
  	   &\beta  \leq a_k\leq \ln\left(\frac{P_\text{r,max}}{\alpha\delta_k}\right), \quad k =2,\cdots,K_0,  \label{ak_inequal}
  	\\
    &\beta  \leq \nu_k \leq 0, \ P_k \geq 0,\  \forall k \in \{1,\cdots, K_0\},  \label{vari_ineq}
   \end{align}
   Here, \eqref{non_convex1}, \eqref{non_convex2}, \eqref{a1_inequal}, and \eqref{ak_inequal} represent the log-forms of constraints \eqref{Ak_upper_ineq}, \eqref{Pk_inequal}, \eqref{A1_inequal}, and \eqref{Ak_inequal2}, respectively, and \eqref{vari_ineq} is obtained by \eqref{Pk_inequal} and \eqref{new_vari_eq}. However,  Problem ($ \text{P}_4 $) is also non-convex since the left-hand sides of inequalities \eqref{non_convex1} and \eqref{non_convex2} are both sums of concave and convex functions.
 \subsubsection{Constraint approximations}    
 According to the basic idea of the SPCA method \cite{Beck},   by substituting the concave functions in \eqref{non_convex1} and \eqref{non_convex2} with their first-order Taylor approximations, Problem ($ \text{P}_4 $) can be solved by iteratively solving a sequence of the following convex problems 
\begin{eqnarray}
   	(\text{P}_4^i)    &\max\limits_{ \left\{\substack{ \nu_1, \cdots,\nu_{K_0} \\ a_1,\cdots,a_{K_0} \\ P_1,\cdots,P_{K_0} }\right\}} &\sum_{k=1}^{K_0}T_kC_{\text{approx}}(P_{k},B_k) \label{cccp_obj}  \\
  	&\text{s.t.} &L_1( \nu_{k-1}, a_{k-1}, \nu^{i-1}_{k-1}, a^{i-1}_{k-1})  + e^{a_k} \leq 0 , \quad k =2,\cdots,K_0,\label{cccp_con1} \\
  	 &\  &L_2( \nu_k, a_k, \nu^{i-1}_k, a^{i-1}_k)+ \frac{\alpha\delta_{k-1}}{2}e^{\nu_k+a_k}  +P_k  \leq 0   , \quad k =1,\cdots,K_0, \label{cccp_con2} \\
  &\   &\eqref{a1_inequal}-\eqref{vari_ineq},  \label{cccp_con4}
   \end{eqnarray} 
   where $ L_1( \nu_{k-1}, a_{k-1}, \nu^{i-1}_{k-1}, a^{i-1}_{k-1}) $ and  $ L_2( \nu_k, a_k, \nu^{i-1}_k, a^{i-1}_k) $ are the first-order Taylor  approximations for the concave functions in \eqref{non_convex1} and \eqref{non_convex2}, respectively,  and they are given as 
  \begin{align}
 	L_1&( \nu_{k-1}, a_{k-1}, \nu^{i-1}_{k-1}, a^{i-1}_{k-1}) \notag \\  &=  - e^{2g_0(f_{k-1})l}(1-\alpha)\delta_{k-1}^2e^{2\nu^{i-1}_{k-1}+a^{i-1}_{k-1}}\left(2\left(\nu_{k-1} -\nu^{i-1}_{k-1}\right)+ a_{k-1} -a_{k-1}^{i-1} +1   \right), \label{cccp_con5}
 \end{align}
 and
 \begin{align}
 	L_2&( \nu_k, a_k, \nu^{i-1}_k, a^{i-1}_k) \notag \\ &= -\frac{\alpha\delta_{k-1}}{4}e^{2\nu^{i-1}_k+a_k^{i-1}}\left(2\left(\nu_k -\nu^{i-1}_k\right)+ a_k -a_k^{i-1} +1 \right) - \frac{\alpha\delta_{k-1}}{4}e^{a_k^{i-1}}\left(a_k -a_k^{i-1} +1\right), \label{cccp_con6}
 \end{align}  
 respectively. Here, Problem ($ \text{P}_4^{i} $) denotes the $ i $-th convex problem for Problem ($ \text{P}_4 $). Besides, $ [ \nu_1^{i-1}$ $, \cdots,\nu_{K_0}^{i-1}, a_1^{i-1},\cdots,a_{K_0}^{i-1} ] $ is denoted as the optimal solution for Problem ($ \text{P}_4^{i-1} $), which can be computed by some optimization tools, e.g., CVX\cite{boyd2004}. 
 
 Now, we summarize the SPCA method in Algorithm \ref{al_spca}.  In this algorithm, input parameters $ K_0 $, $ P_\text{t} $, $ \{f_1, \cdots,f_{K_0-1} \}$, $ \{B_2, \cdots, B_{K_0} \}$, and $ \{\delta_0, \cdots,\delta_{K_0} \}$ have already been computed by Algorithm \ref{al_K_0}. Then,  we iteratively solve convex Problem ($ \text{P}_4^{i} $) via some optimization tools until its solution $ [\nu_1^i, \cdots, \nu_{K_0}^i, a_1^i, \cdots, a_{K_0}^i,  P_1^i, \cdots, P_{K_0}^i] $ converges.  Finally, an approximate solution $[\mu_1^\star, \cdots, \mu_{K_0}^\star, A_1^\star, \cdots, A_{K_0}^\star]$ for original Problem ($ \text{P}_1 $) is obtained by \eqref{new_vari_eq}. Moreover, we denote $\Omega^{\star}$ as the energy efficiency of the considered mobile RBCom system, which is defined as the ratio of maximum throughput given in \eqref{obj} to the product of pumping power $ P_\text{in} $ and communication time $ T_\text{up}^{K_0} $, i.e., 
 \begin{equation}
 	\Omega^{\star}=\frac{\sum_{k=1}^{K_0} T_kC_{\text{approx}}(P_{k}^{\star},B_k)}{P_\text{in}T_\text{up}^{K_0}}, \label{eq_effi}
 \end{equation}
 where $ P_{k}^{\star} $ is computed by \eqref{P_peak}. Here, $ \mu_k $ and $ A_k $ in \eqref{P_peak} are replaced with $ \mu_k^\star $ and $ A_k^\star $, respectively.
    \begin{algorithm}[htbp]
\label{alg_3}
	\caption{SPCA algorithm to solve Problem ($\text{P}_1 $) \label{al_spca} }
	\label{Joint Parameter Estimation And Data Detection} 
	\begin{algorithmic}[1]
	\Require $ \alpha $, $ K_0 $, $ P_\text{t} $, $ \{f_0, \cdots,f_{K_0-1} \}$, $ \{B_1, \cdots,B_{K_0} \}$, $ \{\delta_0, \cdots,\delta_{K_0} \}$, and  $ P_\text{r, max} $.
	\Ensure $[\mu_1^\star, \cdots, \mu_{K_0}^\star, A_1^\star, \cdots, A_{K_0}^\star]$.
    \State Compute $ T_k $ by \eqref{T_k} for $k =1,\cdots, K_0$;
    \State Set $ i = 0 $, $ \beta = -10^3$, and threshold  $ \epsilon =0.01 $;
    \State Randomly choose a starting point $ [\nu_1^0, \nu_2^0, \cdots, \nu_{K_0}^0, a_1^0, a_2^0, \cdots, a_{K_0}^0 ] $ for Problem ($ \text{P}_4 $);
    \State \textbf{Repeat}
    \State \quad \ \  Set $ i = i+1 $;
    \State \quad \ \ Utilize  $ [\nu_1^{i-1}, \cdots, \nu_{K_0}^{i-1}, a_1^{i-1}, \cdots, a_{K_0}^{i-1} ] $ to obtain the $i$-th convex Problem ($\text{P}_4^i$), as shown in \eqref{cccp_obj}-\eqref{cccp_con6};
    \State \quad \ \ Compute solution $ [\nu_1^i, \cdots, \nu_{K_0}^i, a_1^i, \cdots, a_{K_0}^i,  P_1^i, \cdots, P_{K_0}^i] $ for Problem ($\text{P}_4^i$) via optimization tools, e.g., CVX \cite{boyd2004}; 
    \State \textbf{Until} {$ \sqrt{\sum_{k=1}^{K_0}( \nu_k^{i} - \nu_k^{i-1} )^2 + ( a_k^{i} - a_k^{i-1} )^2 + ( P_k^{i} - P_k^{i-1} )^2 } < \epsilon $} 
    \State Obtain solution $[\mu_1^\star, \cdots, \mu_{K_0}^\star, A_1^\star, \cdots, A_{K_0}^\star]$ for  Problem ($ \text{P}_1 $) by \eqref{new_vari_eq};
    \State Compute energy efficiency $ \Omega^{\star} $ by \eqref{eq_effi}, where  $ P_{k}^{\star} $ in \eqref{eq_effi} is obtained by \eqref{P_peak}.
	\end{algorithmic}
\end{algorithm}

 \section{Simulation Results}
In this section, we give some simulation results to validate our theoretical results. Here, we adopt an Nd:YAG rod as the gain medium. Besides, wavelength of the resonant beam is set as $ \lambda = 1064 $ nm, splitting ratio $ \alpha $ is selected to be $ 0.01 $, the effective area of the receiving surface is fixed at $ S = 0.1256 $ $  \text{m}^2$, and diffraction angle is fixed at $ \phi = 0.2 $ mrad. Moreover, parameters  $ f_0$, $ I_\text{s}(f_0)$, $ \Delta f_\text{H}$, $ S_\text{g}$, and $\eta$ are set the same as those in Remark \ref{Re_h_k}.  Channel bandwidth at frame $1$ is set as $ B_1 = 1 $ GHz, and accordingly, the number $N$ of transmitted symbols per frame is computed by \eqref{B_k}. PSD of the AWGN noise is fixed at $ n_0 =  -174 \ \text{dBm/Hz}  $ and the minimum data rate is set as $ C_{\text{th}} = 0.1$ $ \text{bits/s}$.   
\begin{figure}[htbp] 
\centering
\subfigure[ \label{fig_K0_a} Pumping power $ P_\text{in}=200$ W, $\Vert \vec{q}_0\Vert$= $1$ km]{
\includegraphics[width=2.8in]{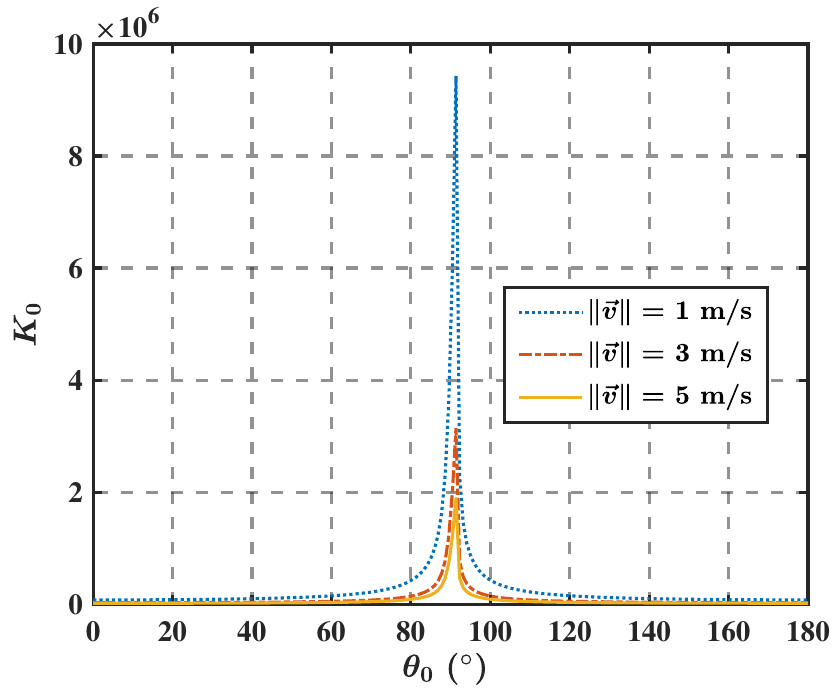}}
 \hspace{0.51in}
\subfigure[ \label{fig_K0_b} Direction angle $ \theta_0= 0^{\circ},\Vert \vec{q}_0\Vert$= $1$ km]{
\includegraphics[width=2.7in]{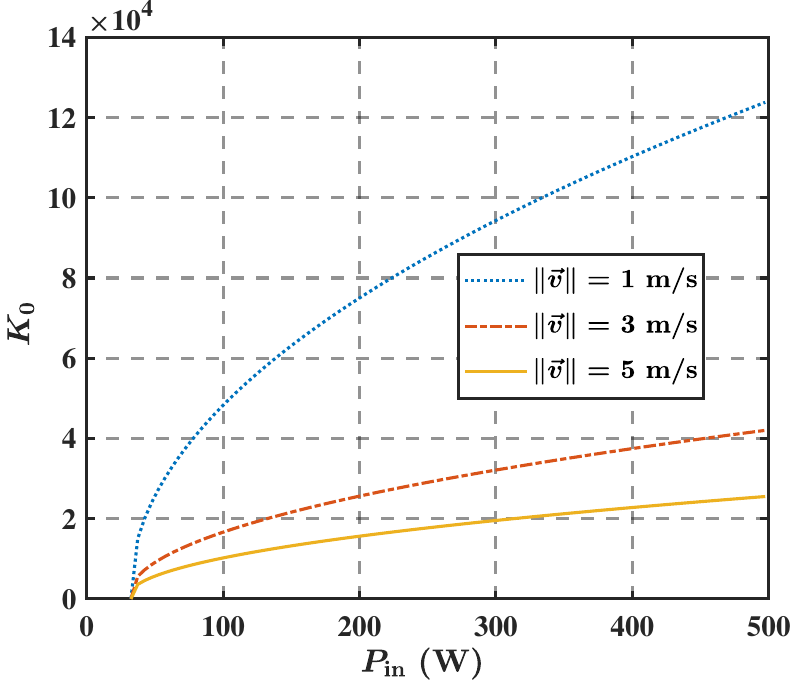}
}
\caption{ Upper bound $K_0 $ on the number of successfully transmitted frames.  }
\label{fig_K0}
\end{figure}

Fig. \ref{fig_K0} describes the variations of upper bound $ K_0 $ on the number of successfully transmitted frames with the initial distance between the transmitter and receiver being $ \Vert \vec{q}_0\Vert$= $1$ km. Specifically, Fig. \ref{fig_K0_a} plots $ K_0 $ as a function of direction angle $\theta_0$, with fixed pumping power of $200$ W. Our observations reveal that $ K_0 $  decreases significantly as $\theta_0$ approaches $0^{\circ}$ or $180^{\circ}$, whereas it attains an extremely large value when $\theta_0$ is in close proximity to $90^{\circ}$. This behavior is attributed to the slow increase in the cumulative Doppler shift in \eqref{f_k}  as $\theta_0$ approaches $90^{\circ}$, resulting in the reduced link gain and increase in $ K_0 $.
Furthermore, we examine the relationship between $ K_0 $ and speed $\Vert \vec{v} \Vert$ of the receiver, and find that $ K_0 $ increases as $\Vert \vec{v} \Vert$ decreases in Fig. \ref{fig_K0_a}. The reason is that the cumulative Doppler shift in \eqref{f_k} increases slower as  $ \Vert \vec{v} \Vert  $ becomes smaller.
 Fig. \ref{fig_K0_b} shows that $ K_0 $ monotonically increases as pumping power $P_\text{in}$ surpasses threshold power $ P_\text{th} $, with direction angle $ \theta_0 $ being $ 0^{\circ} $. However, the rate of increase in $ K_0 $ becomes progressively slower as $P_\text{in}$ increases.

\begin{figure}[htbp] 
\centering

\subfigure[ \label{fig_distance_a} Pumping power $ P_\text{in}=200$ W, $\Vert \vec{q}_0\Vert$= $1$ km]{
\includegraphics[width=2.7in]{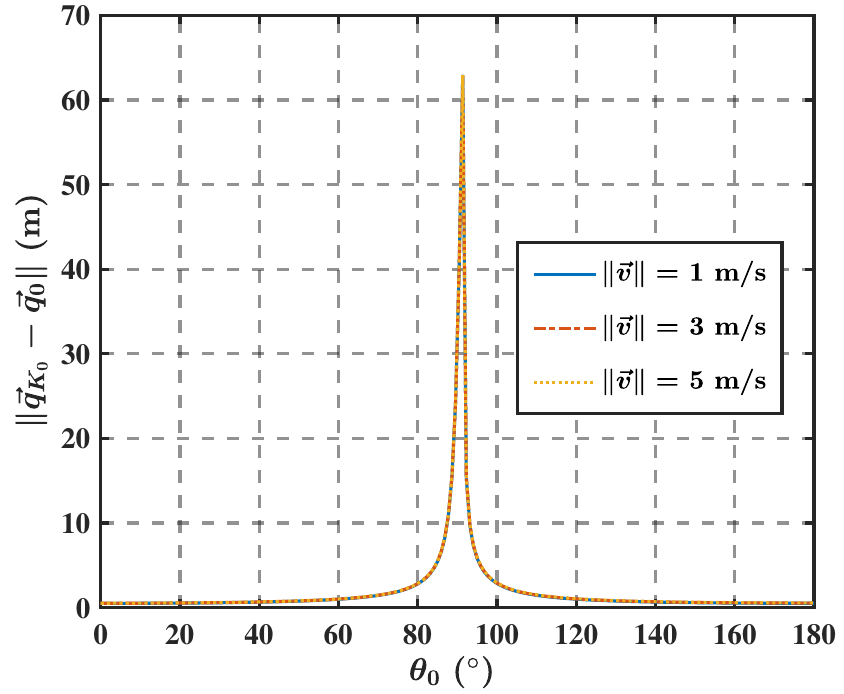}}
 \hspace{0.51in}
\subfigure[ \label{fig_distance_b}  Direction angle $ \theta_0= 0^{\circ},\Vert \vec{q}_0\Vert$= $1$ km]{
\includegraphics[width=2.7in]{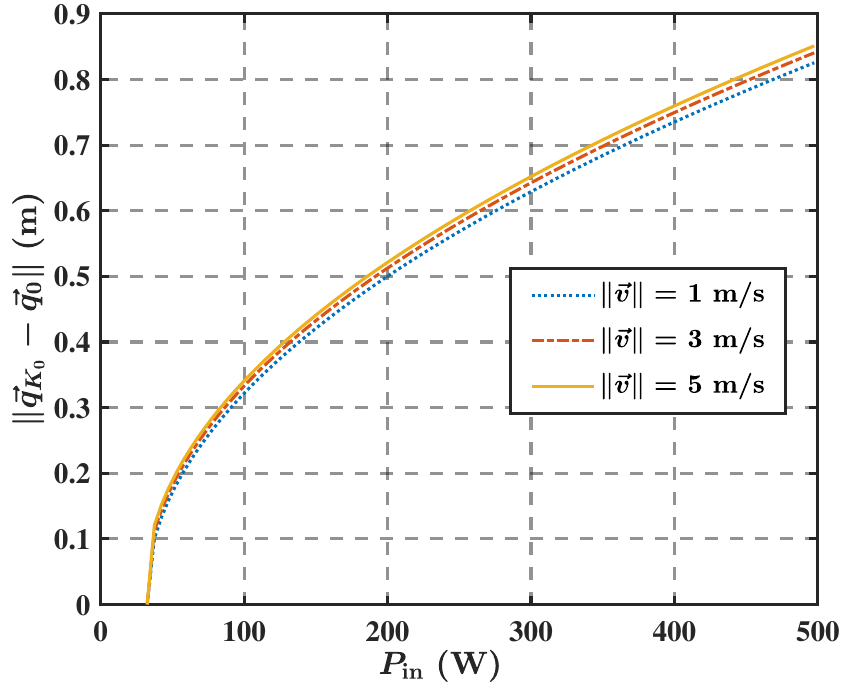}
}
\caption{Upper bound $ \Vert \vec{q}_{K_0} - \vec{q}_0\Vert $ on the moving distance of the receiver.}
\label{fig_distance}
\end{figure}
\begin{figure}[htbp] 
\centering
\subfigure[ \label{fig_time_a} Pumping power $ P_\text{in}=200$ W, $\Vert \vec{q}_0\Vert$= $1$ km]{
\includegraphics[width=2.8in]{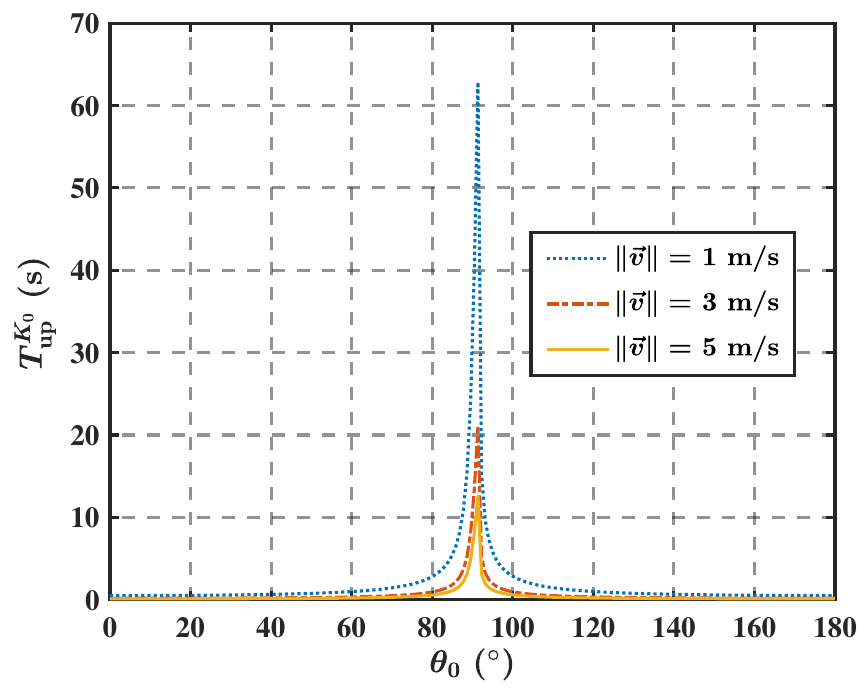}}
 \hspace{0.51in}
\subfigure[\label{fig_time_b}  Direction angle $ \theta_0= 0^{\circ},\Vert \vec{q}_0\Vert$= $1$ km]{
\includegraphics[width=2.7in]{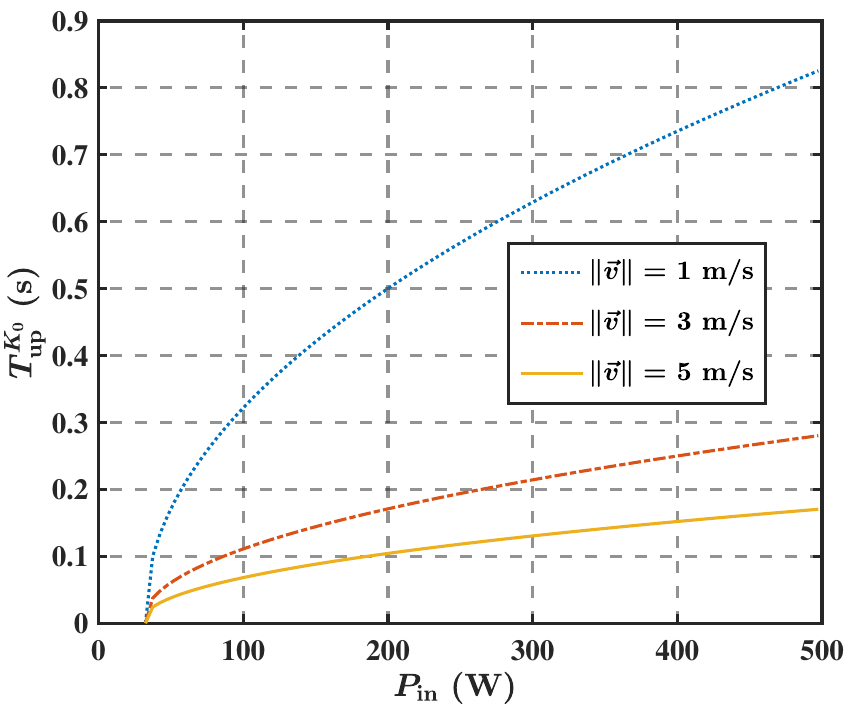}
}
\caption{Upper bound $ T_\text{up}^{K_0} $ of communication time.}
\end{figure}

Then, Fig. \ref{fig_distance} presents the variations of the upper bound $ \Vert \vec{q}_{K_0} - \vec{q}_0\Vert $ on the moving distance of the receiver, with $ \Vert \vec{q}_{K_0} - \vec{q}_0\Vert $ being computed in Proposition \ref{P_qk}. Specifically, compared to Fig. \ref{fig_K0_a},  Fig. \ref{fig_distance_a} shows that $ \Vert \vec{q}_{K_0} - \vec{q}_0\Vert $ is also extremely large when direction angle $\theta_0$ is in close proximity to $90^{\circ}$, and decreases sharply when $\theta_0$ approaches $0^{\circ}$ or $180^{\circ}$. Besides, Fig. \ref{fig_distance_b} shows that $ \Vert \vec{q}_{K_0} - \vec{q}_0\Vert $ monotonically increases with respect to pumping power $ P_{\text{in}} $.   However, both of the two figures in Fig. \ref{fig_distance} show that speed  $ \Vert \vec{v} \Vert $ almost does not affect on $ \Vert \vec{q}_{K_0} - \vec{q}_0\Vert $. After that, as presented in Fig. \ref{fig_time_a} and Fig. \ref{fig_time_b}, we investigate the upper bound $ T_\text{up}^{K_0} $ of communication time in the considered mobile RBCom. It is easy to see that $ T_\text{up}^{K_0} $ is approximately inversely proportional to the speed $\Vert \vec{v} \Vert$ of the mobile receiver increases, resulting in a reduction in $ T_\text{up}^{K_0} $ as $\Vert \vec{v} \Vert$ increases.

\begin{figure}[htbp] 
\centering
\subfigure[ \label{fig_L_K0} $ K_0 $ vs. pumping power $P_\text{in} $]{
\includegraphics[width=2.7in]{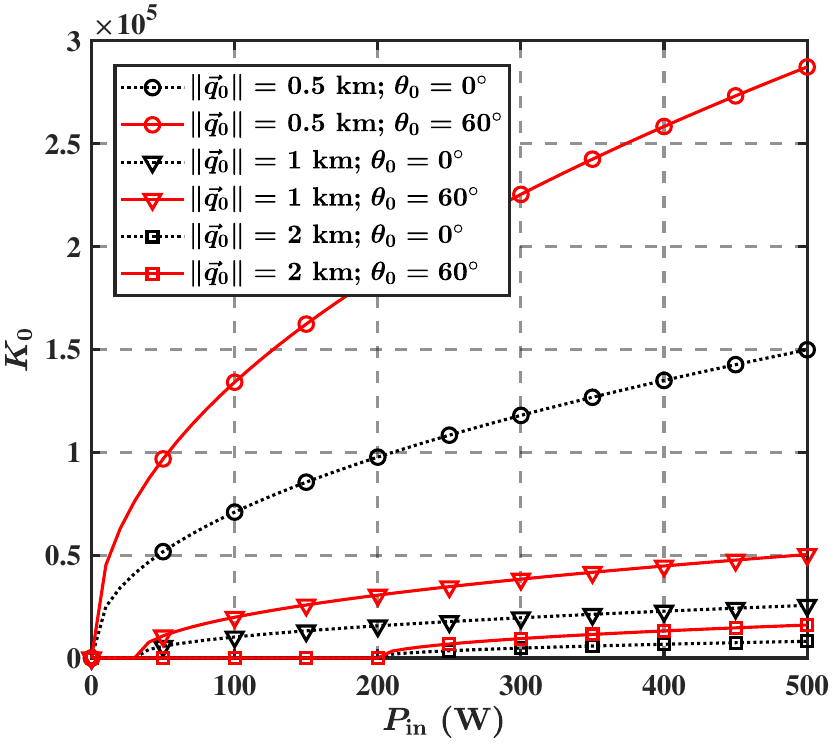}}
 \hspace{0.51in}
\subfigure[ \label{fig_L_distance} $ \Vert \vec{q}_{K_0} - \vec{q}_0\Vert $ vs. pumping power $P_\text{in} $ ]{
\includegraphics[width=2.7in]{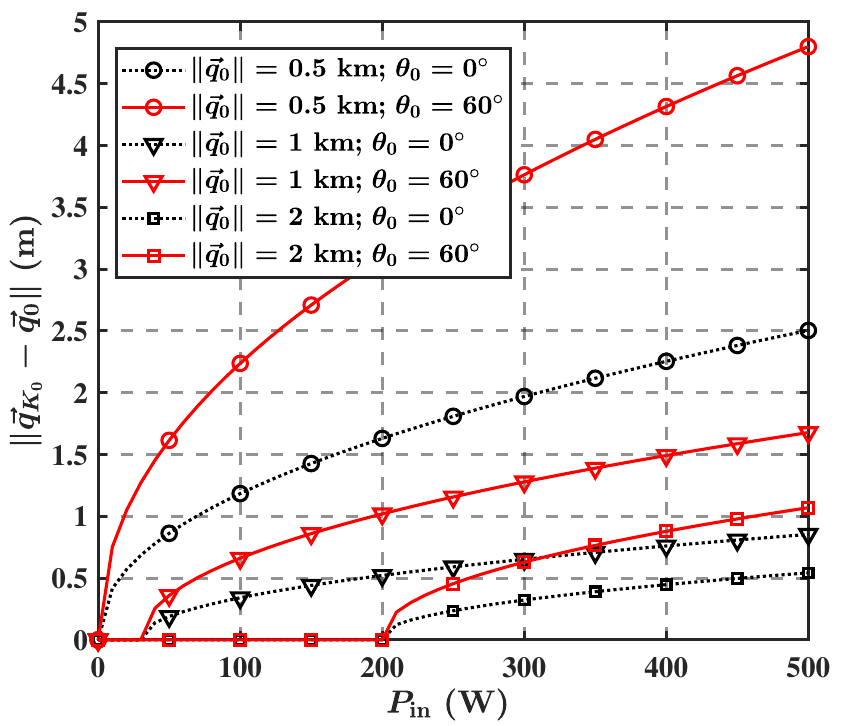}
}
\caption{ Upper bound $ K_0 $ and $ \Vert \vec{q}_{K_0} - \vec{q}_0\Vert $ for different initial distance $ \Vert \vec{q}_0 \Vert $,  direction angle $ \theta_0 $ and  pumping power $ P_\text{in} $ with fixed moving speed $\Vert \vec{v}\Vert$= 5 m/s.}
\label{fig_L}
\end{figure}

Fig. \ref{fig_L}  illustrates variations of  upper bound $ K_0 $ and $ \Vert \vec{q}_{K_0} - \vec{q}_0\Vert $ for different initial distances $ \Vert \vec{q}_0 \Vert $ between the transmitter and the receiver. Fig. \ref{fig_L_K0} reveals that $K_0$ monotonically decreases with an increase in $\Vert \vec{q}_0 \Vert$, owing to the fact that the link loss $\delta_k$  monotonically decreases as $\Vert \vec{q}_0 \Vert$ increases. Additionally, Fig. \ref{fig_L_distance} shows that $ \Vert \vec{q}_{K_0} - \vec{q}_0\Vert $ also monotonically decreases with an increase in $\Vert \vec{q}_0 \Vert$ for fixed $\theta_0$.

\begin{figure}[htbp]
	\centering
	\includegraphics[width=3.5in]{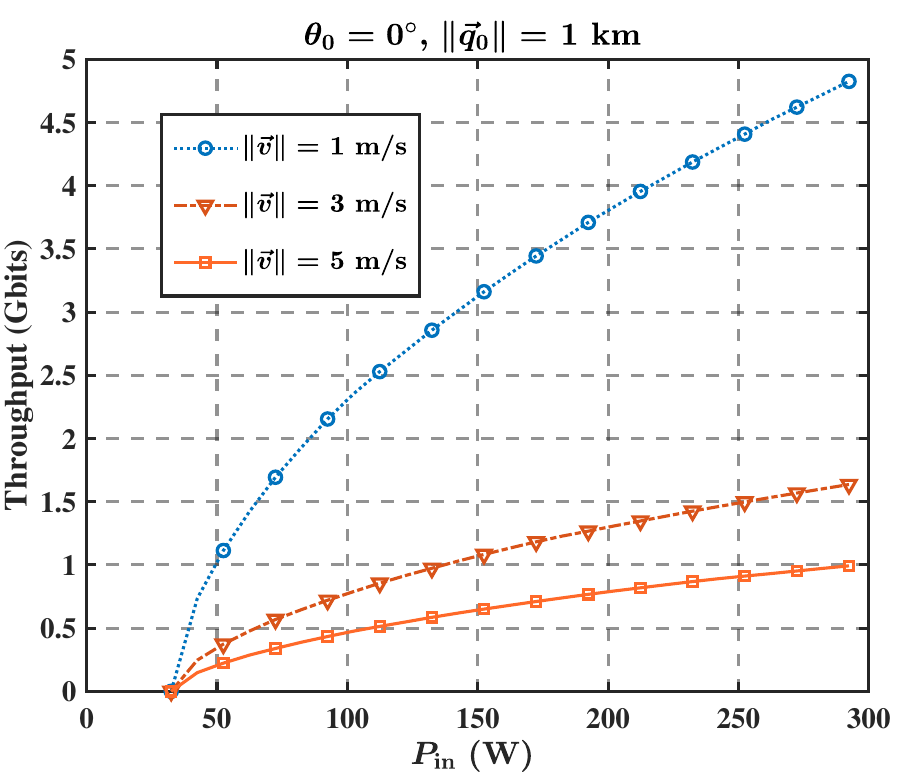}
	\caption{Throughput  vs. pumping power $ P_{\text{in}} $. }
	\label{fig_throughput}
\end{figure} 

\begin{figure}[htbp]
	\centering
	\includegraphics[width=3.5in]{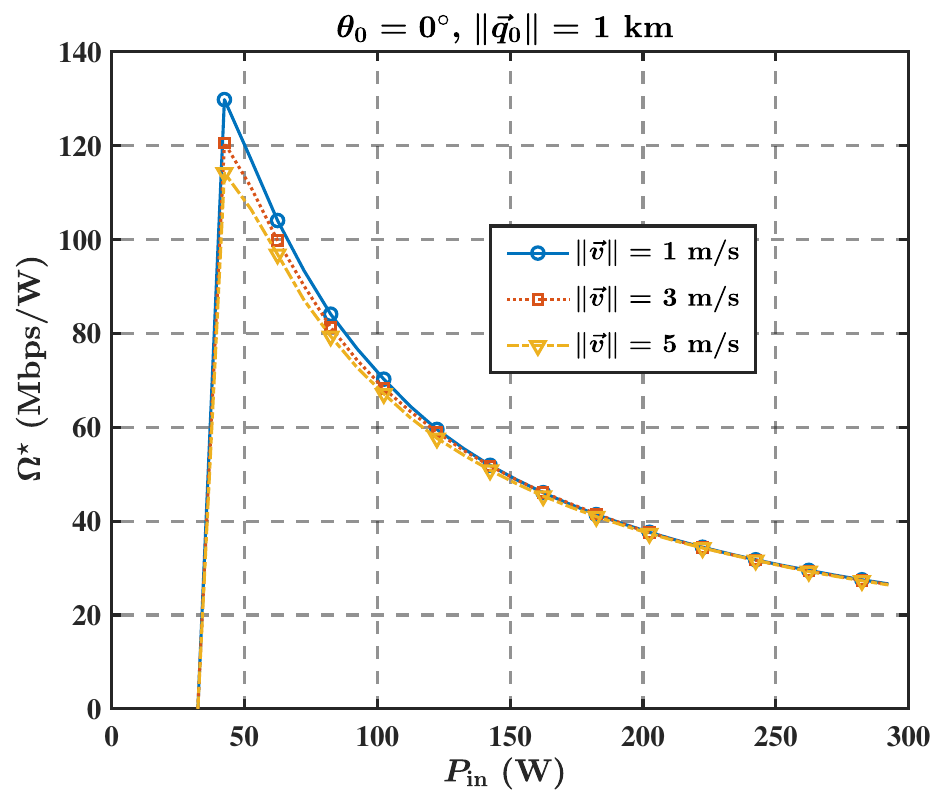}
	\caption{Energy efficiency $ \Omega^{\star} $ vs. pumping power $ P_{\text{in}} $.}
	\label{fig_effi}
\end{figure} 

In Fig. \ref{fig_throughput}, we present the simulation results of the throughput with respect to pumping power $P_{\text{in}}$, for some fixed speed $ \Vert \vec{v} \Vert $, while the initial direction angle is set to $ \theta_0 = 0^{\circ}$. Our observations reveal that, under the given conditions, throughput increases monotonically as $P_{\text{in}}$ increases, and decreases monotonically as the speed $\Vert \vec{v} \Vert$ increases, consistent with the trend exhibited by the communication time in Fig. \ref{fig_time_b}. 

\begin{figure}[htbp]
	\centering
	\includegraphics[width=3.5in]{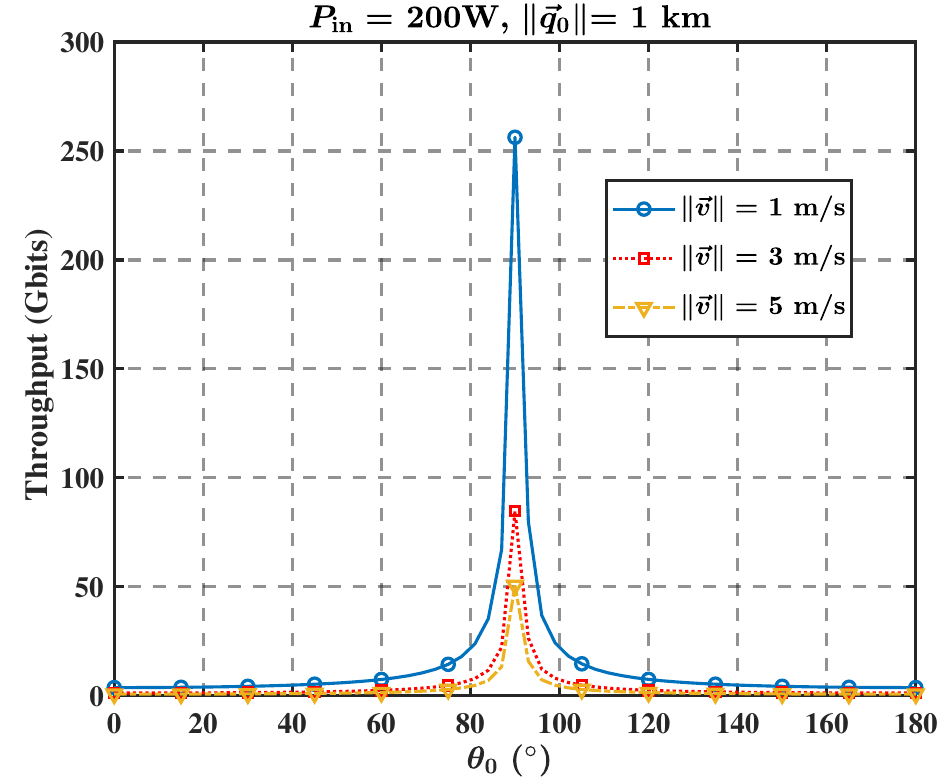}
	\caption{ Throughput vs. direction angle $ \theta_0$. }
	\label{fig_throughput_theta}
\end{figure} 

Then, we show the variations of energy efficiency $\Omega^{\star}$ in terms of the pumping power $ P_\text{in} $, with the parameters being fixed at $ \theta_0 = 0^{\circ} $ and $ \Vert \vec{q}_0 \Vert = 1 \ \text{km} $. As shown in Fig. \ref{fig_effi}, energy efficiency $ \Omega^{\star} $  increases sharply when $ P_\text{in} $ slightly exceeds the threshold pumping power, and then monotonically decreases with increasing $ P_\text{in} $. Moreover, when $ P_\text{in} $ is less than $ 100 $ W, $ \Omega^{\star} $  increases as speed $\Vert \vec{v} \Vert$ increases. However, speed $\Vert \vec{v} \Vert$  has almost no impact on the energy efficiency when $ P_\text{in} $ is larger than $ 100 $ W.

\begin{figure}[htbp]
	\centering
	\includegraphics[width=3.5in]{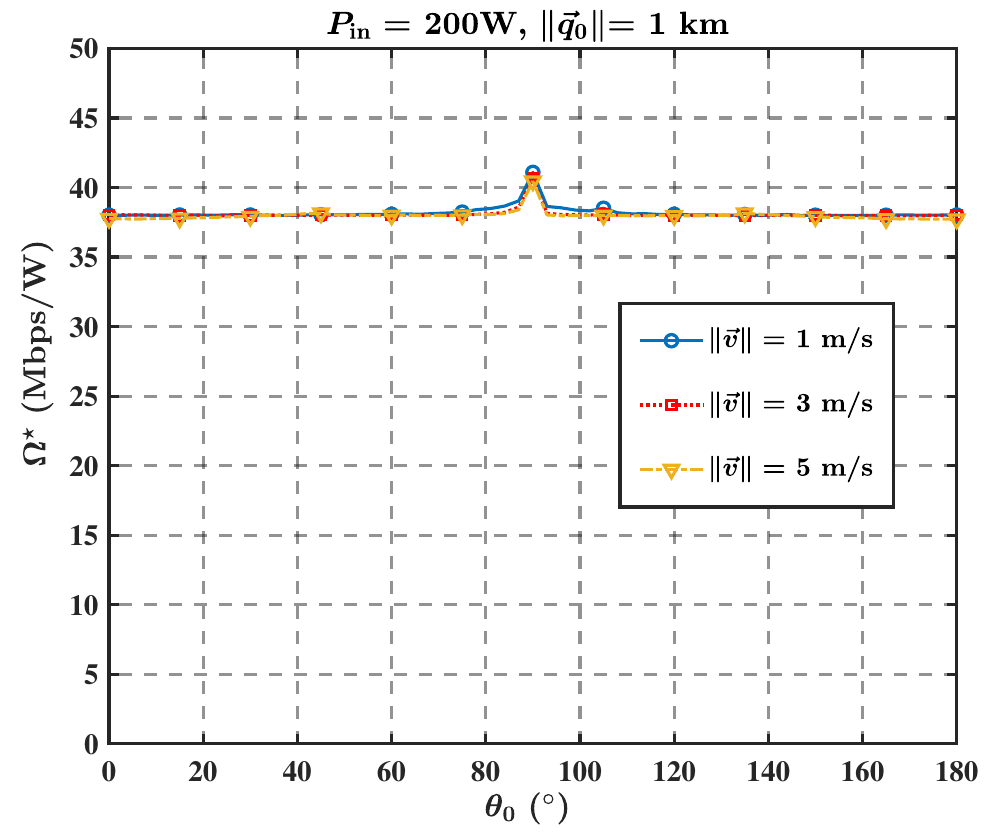}
	\caption{Energy efficiency $ \Omega^{\star} $ vs. direction angle $ \theta_0$. }
	\label{fig_effi_theta}
\end{figure} 

As depicted in Fig. \ref{fig_throughput_theta}, we present the effects of direction angle $ \theta_0 $ on the throughput.  Similar to the communication time performance shown in  Fig. \ref{fig_time_b}, the throughput is also extremely huge when direction angle $ \theta_0 $ is close to $ 90^{\circ} $ and decreases dramatically when $ \theta_0 $ goes to $ 0^{\circ} $ or $ 180^{\circ} $. Additionally, we observe from Fig. \ref{fig_effi_theta} that the energy efficiency $ \Omega^{\star} $  remains almost constant for any direction angle $ \theta_0 $ and different speed $\Vert \vec{v} \Vert$.

\section{Conclusion}

In this paper, we built a new symbol transmission model for the mobile RBCom where the receiver is in relative motion to the transmitter. We investigated the issues of echo interference and the decreasing property of the link gain function in mobile RBCom, and we revealed that the accumulation of the Doppler shift terminates mobile communications. To effectively analyze the communication performance of the considered mobile RBCom, we proposed a simplified channel model by designing a new information-bearing scheme for the modulated symbols, and then provided an algorithm to compute an upper bound on the number of successfully transmitted frames. We also discussed a Doppler shift compensation method to avoid the termination of mobile RBCom. Finally, leveraging the obtained upper bound on the number of transmitted frames, we formulated the throughput maximization problem for the considered mobile RBCom and proposed an SPCA algorithm to effectively solve it. We validated the performance of the proposed method with simulation results in some typical mobile scenarios.

      \appendices
   \section{Proof of Proposition \ref{P_qk}}
   \label{ap_qk}
   As shown in Fig. \ref{fig_move_1}, the receiver moves from $ Q_{k-1} $ to $ Q_{k} $ in the $k$-th reflection round,. Therefore, the position of  $ Q_{k} $ is compute as  $ \vec{q}_k = \vec{q}_{k-1} + \frac{(\Vert \vec{q}_{k-1}\Vert + \Vert \vec{q}_{k}\Vert) \vec{v}}{c} $. Since the speed of the receiver satisfies $ \Vert \vec{v} \Vert\ll c $, we approximately treat $ \Vert \vec{q}_{k-1}\Vert  \approx \Vert \vec{q}_{k}\Vert $ in the $ k $-th reflection round. Then, \eqref{cal_qk} is obtained. Next, since  $ \theta_{k-1} $ is the angle between $ \vec{q}_{k-1} $ and $ \vec{v} $, \eqref{cos_theta} is derived directly. Moreover, since both $  \vec{q}_{0} $ and $\vec{v} $ are given, we can recursively calculate $  \vec{q}_{k} $ and $ \cos\theta_{k-1} $ for every index $ k $  according to \eqref{cal_qk}  and \eqref{cos_theta}.   
      
   \section{Proof of Proposition \ref{prop_G}}
   \label{ap_G}
   
  According to Part I of this paper \cite{Dong_part1}, power gain $ G(I^{\text{T},\text{in}}_k(n), f_k) $ is uniquely computed by the following equation
  \begin{equation}
  	    \label{ap_G_calculate}
     	I^{\text{T}, \text{in}}_k(n) = \frac{I_\text{s}(f_k)\Big(g_0(f_k)l - \ln \sqrt{G\big(I^{\text{T}, \text{in}}_k(n),f_k\big)} \Big)}{G\big(I^{\text{T}, \text{in}}_k(n),f_k\big)-1},
  \end{equation}
  where  $ f_k $ is the central frequency of the resonant beam, $ I_\text{s}(f_k) $ and $ g_0(f_k) $ are the saturation intensity and gain coefficient, respectively, and $ l $ is the length of the gain medium. Then, according to laser theory \cite{laserbook1}, for the homogeneous line-broadening gain medium, we have $ g_0(f_k)  \propto  U(f_k) \triangleq \frac{2}{\pi\Delta f_{\text{H}}  }\frac{1}{1+[2(f_k-f_0)/\Delta f_{\text{H}}]^2} $.  Therefore, $ g_0(f_k) $ is computed as 
   \begin{equation}
   		g_0(f_k) = \frac{g_0(f_0)}{1+[2(f_k-f_0)/\Delta f_{\text{H}}]^2}.
   \end{equation}
   Then, together with $ g_0(f_0)=\frac{\eta P_\text{in}}{I_\text{s}(f_0)lS_{\text{g}}} $ in \cite{Dong_part1}, we have 
   \begin{equation}
   \label{ap_g_fk}
   	    g_0(f_k) = \frac{\eta P_\text{in}}{(1 + \frac{4(f_k-f_0)^2}{\Delta f_{\text{H}}^2} )I_\text{s}(f_0)lS_{\text{g}}}.
   \end{equation}
   Similarly, since the  saturation intensity  satisfies $ I_\text{s}(f_k) \propto \frac{1}{U(f_k)} $\cite{laserbook1}, $ I_\text{s}(f_k) $ is then computed as
    \begin{equation}
     \label{ap_gain_Is}
     	I_\text{s}(f_k) = I_\text{s}(f_0)\{ 1+[2(f_k-f_0)/\Delta f_{\text{H}}]^2  \}.
   \end{equation} 
   Combing  \eqref{ap_g_fk} and \eqref{ap_gain_Is} with \eqref{ap_G_calculate}, \eqref{g_calculate} is derived accordingly. 
   
   Moreover, according to Part I of this paper \cite{Dong_part1} and \eqref{ap_G_calculate}-\eqref{ap_gain_Is}, it is easily observed that $ G(I^{\text{T},\text{in}}_k(n), f_k) $ satisfies the following properties:
   \begin{enumerate}
   	     	  \item Power gain $G\big(I^{\text{T}, \text{in}}_k(n),f_k\big)$  monotonically increases as input intensity $ I^{\text{T}, \text{in}}_k(n) $  decreases. Besides, we have  $ \lim\limits_{I^{\text{T}, \text{in}}_k(n) \to 0 }G\big(I^{\text{T}, \text{in}}_k(n),f_k\big) = \exp(2g_0(f_k)l) $ and  $ \lim\limits_{I^{\text{T}, \text{in}}_k(n) \to \infty }G\big(I^{\text{T}, \text{in}}_k(n),f_k\big) = 1 $.    Moreover, $G\big(I^{\text{T}, \text{in}}_k(n),f_k\big)$  monotonically decreases as $\vert f_k - f_0 \vert $  increases.\label{prop_I_p1}
     	  \item  Output intensity  $ I^{\text{T}, \text{out}}_k(n) $  monotonically increases as input intensity $ I^{\text{T}, \text{in}}_k(n) $ increases.	
   \end{enumerate}
  
  \section{Proof of Proposition \ref{prop_h_k}} \label{ap_prop_h_k}
  To prove this proposition, we first derive some important properties about $ h(x_k^2(n), f_k, \delta_k )$. 
     \begin{Lemma}
   \label{ap_Le_h_property}	 
 Link gain function $ h(x_k^2(n), f_k, \delta_k ) $ has the following two properties:
    \begin{enumerate}
   	   \item   With the other two parameters being fixed, $ h(x_k^2(n), f_k, \delta_k) $  monotonically increases  as  $ x_k(n) $ increases,  decreases as $ \vert f_k - f_0 \vert $ increases, and  increases as $ \delta_k $ increases.
    	\item $ \frac{\sqrt{h(x_k^2(n),f_k,\delta_k)} }{x_k(n)} $  monotonically decreases as $ x_k(n) $ increases and increases as $ \delta_k $ increases. Moreover, it satisfies
 		   	\begin{equation}
    			\delta_k \sqrt{1-\alpha} < \frac{\sqrt{h(x_k^2(n),f_k, \delta_k)} }{x_k(n)} < e^{\frac{\eta P_\text{in}}{\left( 1+ \frac{4(f_k-f_0)^2}{\Delta f_{\text{H}}^2}\right) I_\text{s}(f_0)S_{\text{g}}   }}\delta_k \sqrt{1-\alpha}. 
    			\label{ap_ineq_h}
    		\end{equation}
    \end{enumerate}
   \end{Lemma}
   \begin{IEEEproof}
   	Property 1) is easily obtained by combining \eqref{func_h} and the properties of power gain function $ G(\cdot) $ given in the proof of Proposition \ref{prop_G}. To prove property 2), we first derive $  \frac{\sqrt{h(x_k^2(n),f_k,\delta_k)} }{x_k(n)}  =  \delta_k \sqrt{(1-\alpha)G\left(\frac{(1-\alpha)\delta_k^2x_k^2(n) }{S_{\text{g}} },f_k\right)} $ by \eqref{func_h}. Since $ G\left(\frac{(1-\alpha)\delta_k^2x_k^2(n) }{S_{\text{g}} },f_k\right) \in [1, \exp(2g_0(f_k)l)] $ shown in Appendix \ref{ap_qk}, together with \eqref{ap_g_fk}, \eqref{ap_ineq_h} is obtained accordingly.
   \end{IEEEproof}

    	From \eqref{x_k} and Lemma \ref{ap_Le_h_property}, we have
  	\begin{align}
  		\frac{x_k(n) }{x_{k-1}(n)}  &= \frac{\sqrt{h(x_{k-1}^2(n),f_{k-1},\delta_{k-1})}m_k(n)}{x_{k-1}(n)} \notag \\
  		     &\leq \delta_{k-1} \sqrt{1-\alpha} \cdot \exp{\left(\frac{\eta P_\text{in}}{\left[ 1+ \frac{4(f_{k-1}-f_0)^2}{\Delta f_{\text{H}}^2}\right] I_\text{s}(f_0)S_{\text{g}}   }\right)}, \label{ap_h_decrease}
  	\end{align}
  	with $ k\geq 2 $. Then,  it is easy to see that $ \frac{x_k(n)}{x_{k-1}(n)}  < \sqrt{1-\alpha} $ if $ | f_{k-1} - f_0 | > \frac{\Delta f_{\text{H}}}{2}\sqrt{-\frac{\eta P_\text{in}}{I_\text{s}(f_0)S_{\text{g}}\ln{\delta_{k-1}}}-1} $. Besides, based on our considered mobile scenario that the receiver moves along a fixed direction, the distance $ \Vert \vec{q}_k\Vert$ between the transmitter and the receiver is monotonically increasing or decreasing if  $k$ is sufficiently large\footnote{In some certain mobile scenarios,  $ \Vert \vec{q}_k\Vert$ first decreases and then decreases as $ k $ increases, e.g., direction angle $ \theta_{0}$ is set as $ \theta_{0} = 89^\circ $ in \eqref{cos_theta}. However, in such a scenario, there always exists a sufficiently large $ k^\prime $ such that $ \Vert \vec{q}_k\Vert$ monotonically increases with respect to $ k $ for $k \geq k^\prime$.}, implying that  $|f_k-f_0|$ monotonically increases for sufficiently large $k$ by \eqref{f_k}. Thus, there exists $ k_0 $ such that  $ \frac{x_k(n)}{x_{k-1}(n)}  < \sqrt{1-\alpha} $ holds for $ k \geq k_0 $, which means that $x_k(n)  $ decreases to $0$  approximately exponentially  for  $ k\geq k_0 $. Finally, by \eqref{x_k}, we can easily derive that $ h(x_{k-1}^2(n),f_{k-1},\delta_{k-1}) $ in \eqref{x_k} decreases to $0$  approximately exponentially, as $ k $ becomes sufficiently large. Therefore, this proposition is proved. 
  
  \section{Proof of Proposition \ref{prop_wk}} \label{ap_P_wk}
  
   For the case of $ k =1 $, it is obvious to see that \eqref{w_k} holds if and only if  $ 0\leq  A_1 \leq \sqrt{P_\text{t}}   $. Then, for the case of $ k\geq 2 $, \eqref{w_k}  holds if and only if $ 0\leq A_k \leq \min \limits_{n\in\{1,\cdots,N\}}\sqrt{h(x_{k-1}^2(n),f_{k-1},\delta_{k-1})}  $ 	since $ w_k(n) $ is no greater than $1$. Therefore, to prove this proposition, we only need to derive 
   \begin{equation}
   	\min \limits_{n\in\{1,\cdots,N\}}\sqrt{h(x_{k-1}^2(n),f_{k-1},\delta_{k-1})} = \sqrt{h(A_{k-1}^2\mu_{k-1}^2 ,f_{k-1},\delta_{k-1})},
   	\label{ap_h_wk}
   \end{equation}
   for $ k\geq 2 $. 
   
   Here, considering that \eqref{w_k} holds for the case of $ k =1 $, we prove \eqref{ap_h_wk} by mathematical induction.
   When $k=2$, combining \eqref{x_k} with \eqref{m_k} and \eqref{w_k}, we have 
   \begin{equation}
   	\min \limits_{n\in\{1,\cdots,N\}}x_{1}(n) = A_{1} \cdot \min \limits_{n\in\{1,\cdots,N\}}s_{1}(n) = A_{1}\mu_{1}.
   \end{equation}
   Due to the monotonically increasing property of $ h(x_{1}^2(n),f_{1},\delta_{1})$ with respect to $ x_{1}(n) $ (shown in Lemma \ref{ap_Le_h_property} of Appendix \ref{ap_prop_h_k}),  $ h(x_{1}^2(n),f_{1},\delta_{1})$ satisfies
   \begin{equation}
   	\min \limits_{n\in\{1,\cdots,N\}}\sqrt{h(x_{1}^2(n),f_{1},\delta_{1})} = \sqrt{h(A_{1}^2\mu_{1}^2 ,f_{1},\delta_{1})}.
   \end{equation}
   Therefore, we have proved that \eqref{ap_h_wk} holds for $ k=2 $.  Next, considering that \eqref{ap_h_wk} holds for  $ k = k_0 $ ($ k_0 \geq 2 $),  it is easy to see that \eqref{w_k} also holds for $ k = k_0 $. Moreover, when $ k = k_0+1 $,  combining \eqref{x_k} with \eqref{m_k} and \eqref{w_k}, we have 
    \begin{equation}
    	\min \limits_{n\in\{1,\cdots,N\}}x_{k-1}(n) = \min \limits_{n\in\{1,\cdots,N\}}x_{k_0}(n)= A_{k_0} \cdot \min \limits_{n\in\{1,\cdots,N\}}s_{k_0}(n) = A_{k_0}\mu_{k_0}.
    \end{equation}
     Similarly, utilizing the monotonically increasing property of $ h(x_{k_0}^2(n),f_{k_0},\delta_{k_0})$ in  Lemma \ref{ap_Le_h_property}, \eqref{ap_h_wk} is proved to be ture for $ k=k_0+1 $ and the inductive step is complete. In conclusion, by the principle of mathematical induction, \eqref{ap_h_wk} holds for $ k\geq 2 $. Therefore, the proof of this proposition is complete.

\section{Proof of Proposition \ref{prop_K0} }
To prove this proposition, we first give a simple lemma about approximation $  C_{\text{approx}}(P_{k}, B_k )$ of the channel capacity.
\label{ap_prop_K0}
    \begin{Lemma}
   \label{le_Ak}
   To maximize $  C_{\text{approx}}(P_{k}, B_k )  $, $ A_k^2 $ in \eqref{P_peak} must satisfy
   	\begin{equation}
   A_k^2=  \begin{cases}
   	     P_\text{t}, &k = 1,  \\
   	     h(A_{k-1}^2\mu_{k-1}^2 ,f_{k-1},\delta_{k-1}), &k =2,\cdots,K.
   \end{cases}
   \label{ap_Ak_equal}
   	\end{equation}
   \end{Lemma}
   \begin{IEEEproof}  
  	 Since $ P_{k} $ in \eqref{P_peak} monotonically increases with respect to $ A_k $ and $ C_{\text{approx}}(P_{k}, B_k ) $ in \eqref{appro_C}  monotonically increases with respect to $  P_{k} $, $ A_k^2 $ must achieve its upper bound in \eqref{Ak_range} to maximize $ C_{\text{approx}}(P_{k}, B_k ) $. Therefore, \eqref{ap_Ak_equal} is obtained and we have completed this proof.
   \end{IEEEproof}

From Definition \ref{def_k}, it is easy to see that $ K_0 $ is an upper bound on the number of the successfully transmitted frames if the maximum of $ C_{\text{approx}}(P_{K_0+1}, B_{K_0+1}) $ is less than $C_{\text{th}} $. Besides, noting that $ B_{K_0+1} $ is determined by the considered mobile scenario shown in Fig. \ref{fig_move_1} and computed by \eqref{B_k}, $C_{\text{approx}}(\cdot) $ is maximized by only maximizing $P_{K_0+1}$ since it is monotonically nondecreasing with respect to $P_{K_0+1}$.  Then, it is easy to see that $P_{K_0+1}$ achieves its maximum $ \frac{\alpha\delta_{K_0+1} A_{K_0+1}^2}{4} $ by setting $ \mu_{K_0+1} = 0 $ in \eqref{P_peak}. Moreover, according to \eqref{ap_Ak_equal} and the monotonically increasing property of $h(A_k^2\mu_{k}^2,f_{k},\delta_{k})$ with respect to $\mu_k$ in Lemma \ref{ap_Le_h_property} of Appendix \ref{ap_prop_h_k},  \eqref{Ak_up} is obtained to compute $ A_{K_0+1}^2 $ in $ \frac{\alpha\delta_{K_0+1} A_{K_0+1}^2}{4} $.       
      
   \bibliographystyle{IEEEtran}   
  \bibliography{model_resonant} 
\end{document}